\documentclass{aa}  

\usepackage{graphicx}
\usepackage{txfonts}
\begin{document} 

\title{Insights into the properties of the Local (Orion) spiral arm. NGC 2302: First results and description of the program
       \thanks{Based on observations collected at the Cerro Tololo Inter-American Observatory and at Las Campanas Observatory}}
        
\author{Edgardo Costa\inst{1}, Andr\'e Moitinho\inst{2}, Mat\'ias Radiszc\inst{1},
Ricardo Mu\~{n}oz\inst{1}, Giovanni Carraro\inst{3,4}, Ruben A. V\'azquez\inst{5}, and Elise Servajean\inst{1}}

\institute{Departamento de Astronom\'ia, Universidad de Chile, Casilla 36-D, Santiago, Chile\\
\email{costa@das.uchile.cl}
\and
SIM/CENTRA, Faculdade de Ciencias de Universidade de Lisboa, Ed. C8, Campo Grande, 1749-016, Lisboa, Portugal
\and
ESO, Alonso de Cordova 3107, 19001, Santiago de Chile, Chile
\and
Dipartimento di Fisica e Astronomia, Universit\`a di Padova, Italy 
\and
Facultad   de Ciencias Astron\'omicas y Geof\'isicas (UNLP), Instituto de Astrof\'isica de La Plata (CONICET, UNLP),
Paseo del Bosque s/n, La Plata, Argentina
                   }
\date{}

 
\abstract
{The
spiral structure of the Milky Way (MW) is highly uncertain and is the subject of much discussion nowadays. Even
the spiral structure close to the Sun and the real nature of the so-called Local or Orion arm are poorly
known.}
{We
present the first result from a program that determines the properties of the Local (Orion) spiral arm
(LOA), together with a full description of the program. In this context we have made a comprehensive
study of the young LOA open cluster NGC~2302, which includes a $UBVRI$ photometric analysis and determination
of its kinematic properties --\ proper motion (PM) and radial velocity (RV) -- and of its orbital parameters.}
{Making
a geometric registration of our ad-hoc first- and second-epoch CCD frames (12-year timeframe), we
determined the mean PM of NGC~2302 relative to the local field of disk stars, and, through a comparison
with the UCAC4 catalog, we transformed this relative PM into an absolute one. Using medium-resolution spectroscopy
of 26 stars in the field of NGC~2302, we derived its mean RV. We determined the cluster's structure, center, and
radius by means of a density analysis of star counts. Photometric diagrams for several color combinations
were built using our data, which allowed  us to identify the stellar populations present in the field
of NGC~2302 and to carry out our photometric membership analysis. Isochrone fits to the photometric diagrams
allowed us to determine the fundamental parameters of NGC~2302, including reddening, distance, and age.
The kinematic data and derived distance allowed us to determine the space motion of NGC~2302. This was done by
adopting a time-independent, axisymmetric, and fully analytic gravitational potential for the MW.}
{We
obtained an absolute PM for NGC~2302 of ($\mu_{\alpha} \cos\delta,\mu_{\delta}) = (-2.09,-2.11)$ mas yr$^{-1}$,
with standard errors of 0.410 and 0.400 mas yr$^{-1}$. The mean RV of NGC~2302 turned out to be 31.2 km sec$^{-1}$
with a standard error of 0.7 km sec$^{-1}$. The density analysis revealed a remarkably spherical concentration of
stars centered at $\alpha_{2000}$ = 06:51:51.820, $\delta_{2000}$ = -07:05:10.68 with a radius of $2.5{\arcmin}$.
Although densely contaminated by field stars, all our photometric diagrams show a recognizable cluster
sequence of bright stars ($V \leq$ 18). The color-color diagrams show the existence of more than one population,
each affected by distinct reddening with the cluster sequence at $E(B-V)$ = 0.23. Isochrone fits displaced for
this reddening and for a distance modulus of $(m-M)_0 = 10.69$ (distance, $d=1.40$ kpc) indicate an age
of $\log(t) = 7.90-8.00$ with a slight tendency toward the younger age. Inspection of the shape of the orbit of NGC~2302 and the resulting
orbital parameters indicate that it is a typical population I object.}
{} 
\titlerunning{The open cluster NGC~2302}
\authorrunning{Costa et al.}


\maketitle
%

\section{Introduction}

Although not free of controversy, most authors agree that the Milky Way (MW) has four major spiral arms:
Scutum-Crux, Carina-Sagittarius, Perseus, and Norma-Cygnus (also known as the Outer arm), and at
least one small arm in the vicinity of the Sun, between those of Carina-Sagittarius and Perseus.
This small arm is often referred to as the Local (Orion) arm (hereafter LOA). This paradigm is illustrated
well by the maps of Russeil (2003), usually considered as an up-to-date description
of the Galactic spiral structure.\

Until recently, the LOA had been studied well in the optical domain only up to a few hundred
parsecs from the Sun, and, probably due to the very complex structure of the MW in the third
Galactic quadrant (3GQ), most studies had concentrated on the first and second Galactic
quadrants.  Originally, the LOA was described as a spur, a sort of inter-arm feature departing
from the Sagittarius arm (or splitting it into two arms) in the first quadrant, close to the 
radio source W51 (Avedisova 1985).  More recently, water maser studies (Xu et al. 2013) have suggested that
both spatially and kinematically, the LOA looks like a major arm. It should be noted, however, that
the data used are limited both in number and in volume (only up to about 2 kpc around the Sun),
and are therefore inconclusive.\

Even though as early as 1979 Moffat et al. had suggested that the LOA enters the 3GQ, based on a study
of early-type
stars and HII regions, and that radio observations had
supported this suggestion for long (Burton 1985, from HI; May et al. 1988 and Murphy \& May 1991, both
from CO), only in the past few years has it been recognized as a well-confined elongated structure
in the 3GQ, thanks to the optical/radio studies of Carraro et al. (2005), Moitinho et al. (2006),
and V\'azquez et al. (2008; hereafter Vaz08). These later works, which are based on the optical multiband $UBVRI$
photometry of young open clusters and field stars, complemented with a radio survey of molecular
clouds observed in CO, have shed new light on the structure of the MW in the 3GQ.\

The main conclusions -- in relation to the LOA -- of the thorough analysis presented in Vaz08 were as follows:
1. Usually considered to be a small structure (a {\it Spur}), the LOA is, in fact, substantial.
It extends out to at least 10 kpc from the Sun, all the way into the 3GQ.
2. The LOA crosses the Perseus arm, and the authors hypothesize that it seems to be
disrupting the latter.
While clarifying various aspects of the morphology of the 3GQ, the findings by Vaz08 raised the
following questions about the nature of the LOA:
1. Is it part of a less dense superposed grand design pattern or is it simply a large bifurcation?
2. Does it rotate with the same angular velocity as the four canonical major arms?
3. Is it a transient feature or will it persist similarly to the major arms? (For a brief general
discussion of this matter, see, e.g., D'Onghia et al. 2013)\

These questions can be addressed by retrieving the past structure and evolution of the LOA. This 
can be achieved by means of a robust sample of open clusters that trace the present location and
extent of the LOA in the 3GQ. Its kinematics, distances, positions, and ages are precisely
known. Integration back in time of the orbits of these clusters over the intervals that correspond
to their ages leads to the location of their birthplaces, hences to determination of the
motion of the LOA (see, e.g., Dias \& Lepine 2005). Knowledge of the kinematics of the LOA up to
great distances in the 3GQ will furthermore help determine
if the clusters and early type field stars that constitute the LOA rotate together (as a
rigid body, thus supporting the idea of a stable structure), or not. On the other hand, positive
detection of streaming motions along the arm could support a density wave origin (e.g., Brand \&
Blitz 1993). Comparison of the motion of the LOA with that of the major arms should also give us
insight into the mechanisms that created the former.\

The aim of the present program is to select the appropriate clusters and collect all the data
required to reach the above goal. To this end, we have selected 29 open clusters from the surveys
cited above. They are young and intermediate-age open clusters ($\sim$4 Myr to $\sim$2 Gyr),
located between 217$^{\circ}$ $<$ l $<$ 260$^{\circ}$, and -5$^{\circ}$ $<$ b $<$ +5$^{\circ}$
and at distances between $\sim$2 - 8  Kpc from the Sun. The distribution of our cluster sample in
the Galactic plane is illustrated in Figure 4 of Vaz08, where black circles depict clusters that
probably belong to the LOA.
In Table 1 we present the available parameters for our cluster sample from Moitinho 2001; hereafter M01):
$l$ and $b$ are the Galactic longitude
and latitude, respectively; $(m-M)$ is the distance modulus; $E(B-V)$ is the foreground reddening;
$Rhc$ is the heliocentric distance; X, Y, and Z are the Galactic Cartesian coordinates; and $Rgc$ is the
Galacto-centric distance - adopting 8.5 kpc for the Galacto-centric distance of the Sun. We note that
part of this data has already been published in a series of papers by our collaborators (see, e.g.,
Moitinho et al. 2006).\

Although the data set presented in Table 1 is in many respects unmatched, we still lack the kinematic
information needed to determine the Galactic orbits of the clusters. Throughout the present survey
we obtain this data: precise absolute proper motions (PMs) and radial velocities (RVs).\

In this first paper we describe the procedure used to obtain the PM and RV of the clusters in detail and present
the first result from our program: a comprehensive study of the LOA cluster NGC 2302. The paper is organized
as follows. In Section 2 we describe the observations and methodology in general terms, and in Section 3 we
present the results obtained for NGC 2302.\

\begin{table*}
        \tabcolsep 0.20truecm
        \caption{Parameters of our cluster sample.}
        \begin{tabular}{lcccccrcccc}
          \hline
            \noalign{\smallskip}
            Name & l & b & $m-M$ & $E(B-V)$ & Rhc & Age & X & Y & Z & Rgc\\
                 & $\degr$ & $\degr$ &  mag  &  mag     & kpc & Myr &kpc&kpc&kpc& kpc\\  
            \noalign{\smallskip}
          \hline
            \noalign{\smallskip}
Czernik29   & 230.81 & +0.94 & 13.10 & 0.48 & 4.17 &  200 & -2.63 & 3.23 &  0.07 & 12.02 \\
Haffner10   & 230.82 & +1.00 & 13.60 & 0.52 & 5.25 & 1300 & -3.32 & 4.07 &  0.09 & 13.00 \\
Haffner16   & 242.09 & +0.47 & 13.00 & 0.15 & 3.98 &   50 & -3.52 & 1.86 &  0.04 & 10.94 \\
Haffner18   & 243.11 & +0.42 & 14.50 & 0.60 & 7.94 &    4 & -3.59 & 7.08 &  0.06 & 15.99 \\
Haffner19   & 243.04 & +0.52 & 13.60 & 0.40 & 5.25 &    4 & -2.38 & 4.68 &  0.05 & 13.39 \\
NGC2302     & 219.28 & -3.10 & 10.69 & 0.23 & 1.40 &   40 & -0.88 & 1.08 & -0.07 &  9.62 \\
NGC2309     & 219.89 & -2.22 & 12.00 & 0.35 & 2.51 &  250 & -1.60 & 1.92 & -0.09 & 10.54 \\
NGC2311     & 217.73 & -0.68 & 11.80 & 0.33 & 2.29 &  400 & -1.40 & 1.81 & -0.02 & 10.40 \\
NGC2335     & 223.62 & -1.26 & 11.26 & 0.43 & 1.78 &   79 & -1.23 & 1.29 & -0.04 &  9.87 \\
NGC2343     & 224.31 & -1.15 & 10.26 & 0.18 & 1.12 &  100 & -0.78 & 0.80 & -0.02 &  9.33 \\
NGC2353     & 224.66 & +0.42 & 10.45 & 0.15 & 1.23 &   79 & -0.86 & 0.87 &  0.01 &  9.41 \\
NGC2367     & 235.63 & -3.85 & 11.54 & 0.05 & 1.40 &    5 & -1.64 & 1.12 & -0.13 &  9.76 \\
NGC2383     & 235.27 & -2.43 & 12.65 & 0.30 & 3.39 &  200 & -1.93 & 2.78 & -0.14 & 11.45 \\
NGC2384     & 235.39 & -2.42 & 12.30 & 0.30 & 2.88 &   13 & -1.64 & 2.37 & -0.12 & 10.99 \\
NGC2401     & 229.67 & +1.85 & 14.00 & 0.36 & 6.30 &   20 & -4.80 & 4.07 &  0.20 & 13.46 \\
NGC2414     & 231.41 & +1.94 & 13.80 & 0.50 & 5.75 &   16 & -4.49 & 3.58 &  0.19 & 12.89 \\
NGC2425     & 231.50 & +3.31 & 12.75 & 0.21 & 3.60 & 2200 & -2.80 & 2.24 &  0.21 & 11.10 \\
NGC2432     & 235.48 & +1.78 & 00.00 & 0.23 & 1.90 &  500 & -1.56 & 1.07 &  0.06 &  9.69 \\
NGC2439     & 246.41 & -4.43 & 13.30 & 0.37 & 4.60 &   10 & -4.20 & 1.83 & -0.35 & 11.16 \\
NGC2453     & 243.35 & -0.93 & 13.60 & 0.50 & 5.25 &   40 & -4.69 & 2.35 & -0.08 & 11.82 \\
NGC2533     & 247.81 & +1.29 & 00.00 & 0.14 & 1.70 &  700 & -1.57 & 0.64 &  0.04 &  9.27 \\
NGC2571     & 249.10 & +3.54 & 10.85 & 0.10 & 1.40 &   50 & -1.30 & 0.50 &  0.08 &  9.09 \\
NGC2588     & 252.29 & +2.45 & 13.47 & 0.30 & 4.95 &  450 & -4.70 & 1.50 &  0.21 & 11.05 \\
NGC2635     & 255.60 & +3.97 & 13.01 & 0.35 & 4.00 &  600 & -3.86 & 0.99 &  0.27 & 10.25 \\
Ruprecht18  & 239.94 & -4.92 & 12.25 & 0.64 & 2.81 &  160 & -2.42 & 1.40 & -0.24 & 10.19 \\
Ruprecht55  & 250.68 & +0.76 & 13.31 & 0.45 & 4.60 &   10 & -4.30 & 1.52 &  0.06 & 10.90 \\
Ruprecht72  & 259.55 & +4.37 & 12.40 & 0.25 & 3.02 & 1200 & -0.55 & 2.96 &  0.23 & 11.48 \\
Ruprecht158 & 259.55 & +4.42 & 13.10 & 0.25 & 4.17 & 1500 & -0.75 & 4.09 &  0.32 & 12.61 \\
            \noalign{\smallskip}
          \hline
        \end{tabular}
      \end{table*}
      
      \section{Observations and methodology}

\subsection{Photometry}

The photometric data used in this survey was secured by Moitinho (2001; hereafter M01) with the
purpose of studying the star formation history and spatial structure of the Canis Major-Puppis-Vela
region. These observations were made with the Cerro Tololo Inter-American Observatory (CTIO)
0.9m telescope. Full details about the photometry exploited here can be found in M01. We note,
however, that the analysis of a fraction of the data described in M01 has not been published.

\subsection{Proper motions}

\subsubsection{Observations}

First-epoch PM imaging of our cluster sample was provided by the M01 survey, which was carried out
between 1994 and 1998. The CTIO 0.9m telescope has been equipped since then with the same imager, making it
very suitable for astrometric programs that require geometrical stability. The CFIM+T2K imager available
on the 0.9m telescope consists of a Tektronic $2048\times 2048$~pix$^2$ CCD detector with 24~$\mu m$ pixels,
which yields a field of view (FOV) of $\sim13.5\arcmin \times 13.5\arcmin$ and a scale of $\sim0.401\arcsec$
per pixel.\

Using the same telescope and set-up, we started to secure second-epoch PM imaging in February 2010.
The $R$ bandpass was chosen for the PM work, in order to minimize the effects of refraction, and for the
same reason, the observations were restricted to less than $\sim$1.5 hr from the meridian. An effort was
made to secure the second-epoch observations of a particular field in pointing conditions similar to
those achieved in the first epoch.\

Both in the first and second PM epochs, four frames were typically obtained for each cluster field,
two shallow and two deep, in order to observe the brightest and faintest stars of interest
with the best possible S/N. Exposure times varied between $\sim$5-15 sec for the shallow frames,
and between $\sim$100-600 sec for the deep frames, depending on each cluster.\

The scale of this set-up, in combination with the generous time base between our first- and second-epoch observations (12.16 years), was deemed sufficient to achieve the required PM precision.
Indeed, in the course of various astrometry programs (Costa, et al. 2011, 2009, 2006, 2005),
we have learned that, for precise relative astrometry, a signal-to-noise
ratio (S/N) of $\sim$150 is required. With this S/N it is possible to measure the X,Y centroids of well-exposed images with a precision
better than $\sim$1/50 of a pixel, using the centering tasks in the
DAOPHOT package (Stetson 1987). Given the scale of our setup, this translates to a positional
precision at any epoch of $\sim$10 mas, for $R$$\leq$19.5. Our expectation is absolutely consistent 
with the positional precision reported by Jao et al. (private comm.) during the 0.9m CTIO parallax
investigation (CTIOPI) program: 2-20 mas, depending on magnitude and exposure time.
With the above positional precision, our expected PM precision for any pair of
observations will therefore be $\sim$0.9-1.2 mas/yr, depending on the time base.\

We have so far obtained excellent PM data for about 70\% of our cluster sample.

\subsubsection{Pixel coordinates}

The coordinates of the stars on each CCD frame were determined using the various routines within
the DAOPHOT package (Stetson 1987). All frames available for each cluster field were first examined to identify the best shallow and best deep image of each PM epoch. Having chosen the best pairs on the basis of image
quality, all objects
in them down to instrumental magnitude limits of roughly 20 in the shallow frame and 22 in the deep frames,
were automatically identified by means of the DAOFIND and PHOT tasks. In this way, preliminary lists
of roughly 2000 and 6000 stars were identified in the selected shallow and deep frames, respectively, of
each cluster field.
The image profiles of the objects in these lists were then examined on an individual basis to discard
objects problematic for astrometry (e.g., too close to a bad CCD column or to the edges, multiple objects
not detected by DAOFIND, galaxies, resolved -- but blended -- objects), which reduced the number of stars
in the lists by about 30\%.\

A subset of typically 180 of the stars in each list was then selected to determine a master PSF for each
frame. For this purpose we used the tasks PSTSELECT and PSF with {\it function = auto} and {\it varoder = 2},
thus allowing the PSF to vary with position on the CCD chip. Experiments carried out to test the centering
parameters of the centering algorithms confirmed that for our purposes the fitting radius
is the most relevant parameter in the PSF fitting process (as was reported in Costa et al. 2009, for
another set-up). Given the conditions in which the reference stars were chosen, the adopted fitting
radius for any frame was always slightly larger than the average FWHM of stellar images. Finally,
by means of the task PEAK, the master PSFs were used to calculate the (X,Y) centroids of all the
stars in the working lists.

\subsubsection{Intra-epoch registration}

A unique list of reference stars for each PM epoch was created by merging the corresponding shallow and
deep lists. To do this, the coordinates of the stars in the shallow frame must be transformed ( registered)
to the system of coordinates
of the deep frame. This was achieved by means of an ad-hoc piece of software that first
identifies the stars in common. Then, using these stars determines the geometrical transformation, and
finally we applied the transformation in the desired sense. Our software
makes use of the IRAF\footnote{IRAF is distributed by the National Optical Astronomy Observatory, which
is operated by the Association of Universities for Research in Astronomy, Inc., under cooperative agreement
with the National Science Foundation} tasks GEOMAP and GEOXYTRAN. For the task GEOMAP, which computes
the spatial transformation function (a polynomial), a general fitting geometry was used that involves shifts,
rotations, scale changes, and higher order optical distortions. Numerous tests were carried out to select the
proper terms and order of the polynomial to be used in each case.
These tests consist of plotting the X and Y residuals of the registration versus the X and Y
coordinates and varying the order of the fits in order to remove all trends in the residuals and
miniminize the rms of the transformation.
Orders as high as six were needed for some terms to remove the optical distortions.
We note that although the higher order optical distortion terms are very small
(the coeffients are comparable to their errors), including them produces a small but
visually noticeable improvement in the fits. It should also be noted that the inclusion,
or not, of the highest order terms does not have a relevant impact on the final results.
GEOXYTRAN simply applies the geometrical transformation (from
"shallow to deep" in our case). The merged list will obviously include repeated entries (objects that were
detected in both frames). Having identified these cases, the lower S/N detections were deleted.

\subsubsection{Inter-epoch registration}

The PM of the clusters is determined by registering first- and second-epoch X,Y coordinates of probable
cluster members and then by applying the geometrical transformation thus derived to all stars in the FOV.
By construction, when this transformation is applied {\it \emph{non}}-cluster members will show higher PM residuals
(formally defined in the next section), whose mean value gives the reflect motion of the field with respect to
the cluster. It should be noted that the internal velocity dispersion of the clusters is much smaller than
that of the field stars and that their mean values generally differ (see, e.g., M\'endez \& van Altena 1996).
Further details are given in the following sections.\

The first step in that direction is to select one of the PM epochs as the "master" epoch. Examination
of all frames available in both epochs showed that, even though seeing was similar in all
observing runs ($\sim$1.0-1.4$\arcsec$), second-epoch frames were of higher quality on average. For
this reason our second PM epoch was chosen as the master epoch, and the first epoch coordinates had to
be transformed to the system of coordinates of the second epoch. This latter system is defined by the X,Y
coordinates of the objects in the merged second-epoch list described above: the {\it \emph{master}} coordinates.\

As explained in Section 2.3 (radial velocities), we are starting to secure medium-resolution spectroscopy
of a significant number of stars in the field of each cluster. From these observations we derive a RV
distribution and select the stars located at the peak of RV histogram as high-probability cluster members.
(This selection is refined further by taking the radius of the cluster into consideration and by using a
photometric membership analysis.) In principle, this subset
of likely cluster members could suffice to derive the geometrical transformation mentioned in the first
paragraph of this section, but many tests showed that in practice their small number (only a few sets of ten)
clearly does not allow dealing with the optical distortions. To solve this problem, a two-step procedure
was required. We first carried out a high-order registration using all stars common to the merged lists
of each epoch and transformed the first epoch coordinates to the system of coordinates of the second
epoch.  Again, numerous tests were carried out to select the proper terms and order of the polynomial.
In this case, orders as high as nine were needed for some terms to remove the optical distortions.
We note that the same considerations for selecting the order of the fit, as described in the previous section,
are valid for the inter-epoch registration.
In a second step, we registered these first-epoch coordinates corrected for distortions into the second-epoch
reference system, this time using only probable cluster members to determine the geometrical transformation
and applying only simple X and Y shifts.\

Although in principle a kinematic membership analysis could be carried out without the aid of a RV
distribution by selecting likely cluster members on the basis of their low residuals (in an iterative
process), in practice this approach could be useful only in the case of our nearest targets. Indeed,
at distances greater than $\sim$4 kpc, our error in the tangential velocity component resulting from
a PM precision of $\sim$1 mas yr$^{-1}$ will be higher than the typical velocity dispersion of the field
(disk) stars ($\sim$20 km/s), making it very difficult to isolate individual cluster members from the field
on the basis of only the PM.

\subsubsection{PM residuals}

PM residuals are defined as the difference between the transformed first-epoch coordinates and the
master coordinates. These residuals allow cluster and field stars to be distinguished, and they lead to the final
determination of the PM of the clusters relative to the field. PM residuals is one of the outcomes of the
second registration described immediately above. The mean value of these residuals for probable cluster
members selected from the RV histogram should be virtually zero, which is indeed the case for our data.
To determine the PM residuals for likely {\it \emph{non}}-cluster members, we made use of a $V$ {\it vs.} $(V-I)$
color-magnitude diagram (CMD) in which we identify stars away from the main cluster sequences. This list of stars is further refined by removing objects with very small PM residuals, suggesting that they
could be cluster members.

\subsubsection{Proper motions}

The PM obtained with the above procedure is the reflected PM of the local field stars with respect to the
cluster; a simple change of signs leads to the PM of the clusters with respect to the field. The transformation
of this relative PM into the system of the International Celestial Reference Frame (ICRF/ICRS, Arias et al. 1995)
is achieved through a comparison with the UCAC4 catalog (Zacharias, 2012) absolute PMs (which are in
the system of the ICRF).

The usual procedure for placing relative PMs on the ICRS is to calculate an average of the differences between the
available relative PM and the ICRS referred PM for common stars, and apply a local correction (Vicente et al.
2010). For this procedure to be meaningful, a healthy number of common stars is required, more than the
high-probability cluster members located on the peak of our RV histogram, and they should also have UCAC4 PMs. For this
reason, we used a variation of this procedure: derive the mean UCAC4 PM of the field stars in a region,
centered on the cluster, which large enough so that the effect of cluster members in the results is irrelevant, and
thus determine the absolute PM of the field, which is applied as a correction to the PM of the cluster relative
to the field. A few tests showed that a $0.5\degr \times 0.5\degr$ zone proved adequate for this purpose.
Objects with high PM and/or high PM errors, or warning flags in the catalog, were deleted. Additionally,
using the APASS $B,V$ photometry available with UCAC4, $V$ {\it vs.} $(B-V)$ CMDs of these zones were
constructed to identify and remove background giant stars. UCAC4 data was downloaded using the VizieR catalog
access tool, the CDS, Strasbourg, France.\footnote{http://vizier.u-strasbg.fr}

\subsubsection{UCAC4 absolute proper motions}

The mean absolute PM of our clusters can be obtained directly from the UCAC4 catalog by averaging the
UCAC4 PM of the list of probable cluster members, selected from the RV histogram, and supplemented by
high-probability cluster members (with no RV determination) in turn selected by means of a photometric membership
analysis.  Again, to calculate the averages, all objects with high PM and/or high PM errors were excluded.
Although for {\it \emph{some}}
of our target fields, the UCAC4 may lack the required angular resolution, precision, or homogeneity, the PMs
thus derived provide a very useful sanity test.

\subsection{Radial velocities}

By means of medium-resolution spectroscopy for a significant number (typically 50 to 100) of stars
in the field of each cluster, the RV distribution, hence the mean RV, of the clusters can be
obtained. Given the typical velocity dispersion in open clusters (less than 5 km/s), we have aimed at
a RV precision of $\sim$2-3 km/s to be able to carry out a membership analysis for all our cluster
sample.

\subsubsection{Observations}

Because of the magnitude range of our targets ($V$$\sim$11-16), the resolution and S/N needed to
achieve the above RV precision can only be obtained with a 4-8m class telescope. In 2011, we started to
observe the less populous clusters ($\sim$50\% of our sample), composed mainly of bright stars
(V$\sim$11-14.5) with the Hydra-CTIO multi-object spectrograph available on the CTIO 4m Blanco telescope.
The Hydra Spectrograph consists of a 400mm Bench Schmidt camera, a SiTe $2048\times4096$pixel CCD
with 15$\mu m$ pixels, and 138 300$\mu m$ (2$\arcsec$) fibers. For efficiency, we used the KPGL3 grating
(527 lines/mm), which provides a resolution of 0.70 \AA/pixel and a coverage of $\sim$3800 \AA. To improve
resolution, a 100$\mu m$ slit plate was used. The SITe CCD was operated in the {\it High S/N} mode; 
2.4 e$^-$/ADU gain, implying a readout noise of 5.2~e$^-$. We expect to target the remaining faint sample
with the ESO/VLT 8m telescope.\

\subsubsection{Basic reductions}

The CCD frames were calibrated using standard IRAF tasks in the CCDPROC package. For this purpose,
Zero and Dome Flat frames were taken every night. After preliminary processing, one-dimensional
spectra were extracted and wavelength-calibrated using the IRAF task DOHYDRA. To this end,
a comparison lamp (PENRAY: He, Ne, Ar) was taken through all fibers at each target pointing.

\subsubsection{Extraction of the radial velocities}

RV were derived by means of the standard cross-correlation technique of Tonry \& Davis (1979), implemented 
in the IRAF FXCOR package. After some testing, it was determined that (for the data we have reduced so far)
the best spectral range to use in the cross-correlation was 5500--6700~\AA\.\

For this purpose, RV standards were observed every night. We note that one observation of a RV standard implies typically
observing it through ten different fibers, which increases the number of spectra available for each
standard by a factor of ten. These observations were supplemented with RV standard spectra obtained in other
RV programs, which used the same set-up.  To check for consistency, each RV standard spectrum was cross-correlated
with every other RV standard spectrum.\

Finally, each target spectrum was cross-correlated against every RV standard spectrum available, producing
multiple RV results for each cluster star observed. After deleting clearly discrepant values, the mean of those
values was adopted as the stellar RV.

\section{First result: NGC 2302}

\subsection{Overview}

Although results for NGC 2302 have been included in a number of large scale studies of open-cluster surveys aimed at a
variety of purposes (see, e.g., Santos-Silva \& Gregorio-Hetem 2012; Kharchenko et al. 2009a,b; de La Fuente Marcos et al. 2009;
Moitinho et al. 2006; Kharchenko et al. 2005; Janes \& Adler 1982), the only previous dedicated observations of NGC~2302 are
those of Moffat \& Vogt (1975), who obtained photoelectric $UBV\beta$ photometry of 16 stars in the cluster field.  To the
best of our knowledge, this is the first deep and comprehensive study of NGC 2302.
Fundamental parameters for NGC 2302 are given in Table 1, except its ICRS equatorial coordinates, which are
$\alpha=06^{\rm h}~51^{\rm m}~.9$, $\delta=-07\degr 05\arcmin$ (J2000)

    \begin{figure*}
     \centering
     \includegraphics[width=\columnwidth]{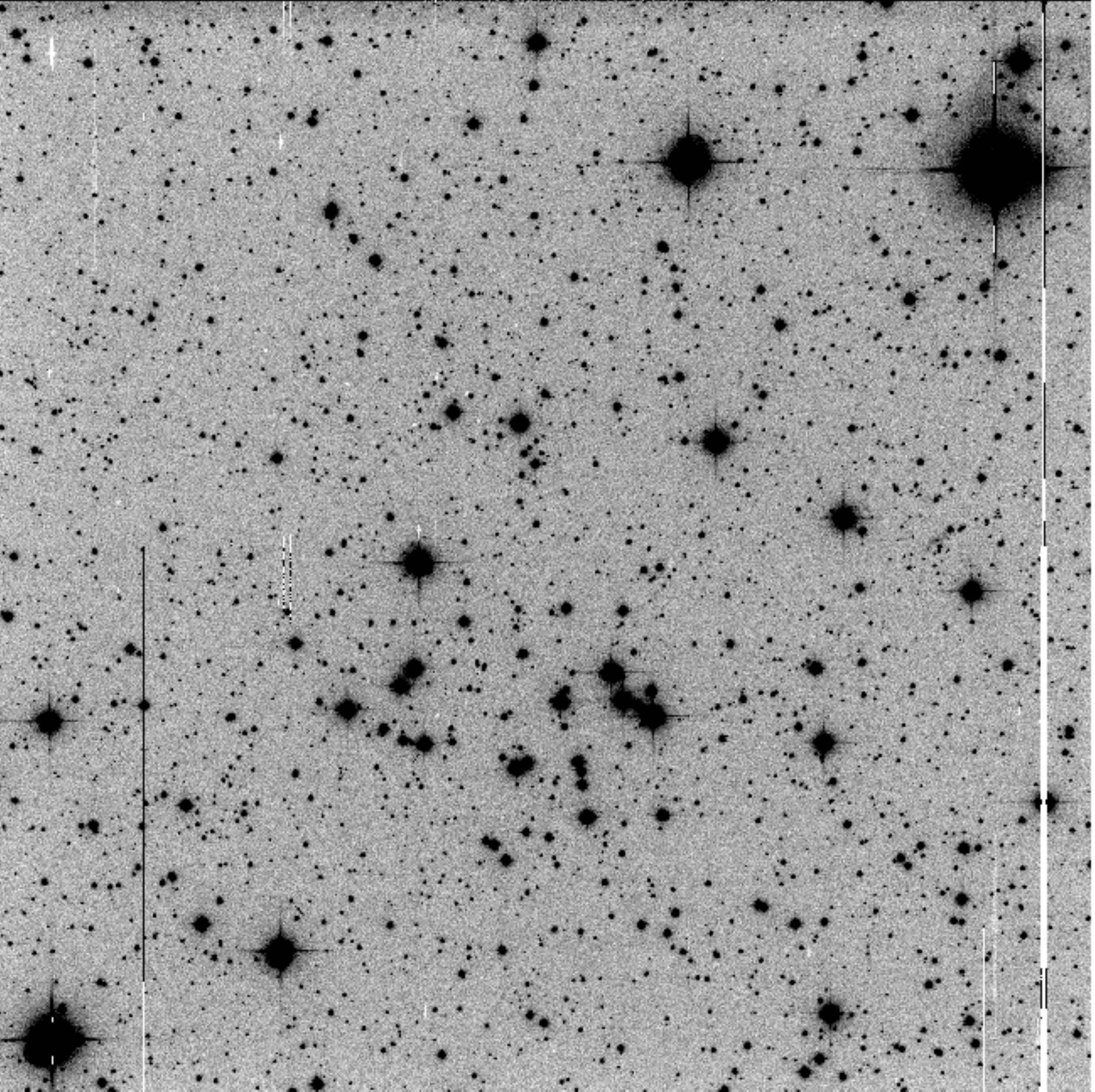}
     \caption{$R$-band image of the NGC 2302 field. The image is $13.5\arcmin$ on a side. North is up, east to the left.
A              120 sec exposure taken with the CTIO 0.9m telescope.}
     \label{ngc2302field}
    \end{figure*}

\subsection{Photometry}

Our photometric study of NGC 2302 is based on the $UBVRI$ data from the M01 survey (see Sections 2.1 and 2.2.1),
and complementary deep $VI$ observations carried out with the Dupont 2.5m telescope at Las Campanas, Chile (LCO).
This latter photometry was secured with a Tektronic $2048\times 2048$~pixel CCD detector, with 24~$\mu m$ pixels,
attached to the Cassegrain focus of the Dupont telescope. The LCO set-up yields a FOV of $\sim8.85\arcmin \times 8.85\arcmin$,
and a scale of $\sim0.259\arcsec$ per pixel.\

\subsubsection{Cluster structure and radius}

From Figure 1, we see that NGC~2302 is revealed as a concentration of bright stars slightly below the center of the image.
In agreement with previous studies from our group, (see, e.g., Moitinho 1997), we address the determinations of the cluster
structure, center, and radius by means of a density analysis of star counts. Here, the density maps have been computed using
the kernel density estimator implemented in the python \emph{scipy.stats.gaussian\_kde} class. The kernel bandwidth was set
by the default \emph{scott} method.\\

A blind density estimation using the whole photometric catalog down to $V$= 18 -- expecting that the cluster over density
would pop out -- does not work in this particular field. The complex reddening structure of the field introduces fluctuations
in the density distribution, which must be carefully considered. The effect of these fluctuations is illustrated in the
lefthand panel of Figure 2, where it can be seen that the highest density is dramatically offset with respect to the cluster.
In contrast, as shown in the righthand panel of Figure 2, if the density map is built using only stars within 0.1 mag of the
cluster's photometric sequence (see Section 3.2.2 and Figure 3), again down to $V$= 18, the cluster over density clearly
emerges, revealing a remarkably round concentration of stars centered at
$\alpha_{2000}$ = 06:51:51.820, $\delta_{2000}$ = -07:05:10.68
($\alpha_{2000}$ = 102.965916$^{ \circ}$, $\delta_{2000}$ = -7.086300$^{\circ}$),
and within a radius of $2.5{\arcmin}$. Beyond this radius, the stellar density reaches the general density of the field.
This limit includes {\it \emph{most}} of the brighter and bluer stars ($V <$ 13 and $B-V < 0.6$) that have visually
led to identifying NGC~2302. We note, however, that a few of those stars that lie outside the derived cluster limit,
though expected to be field stars, have radial velocities and PMs compatible with being cluster members.\\

In any case, the highest density peak seen in the map obtained with all stars down to $V$ = 18 (left panel of Figure 2),
which is centered at 
$\alpha_{2000} \sim$ 103.007810$^{ \circ}$, $\delta_{2000}$ = -7.011069$^{\circ}$, does not include the bright blue stars that
expose the cluster (except for one star). Although this density peak is also seen in the map built using only the stars close
to the cluster's photometric sequence (right panel of Figure 2), it is clearly less pronounced than the cluster's density
distribution.\\

Previous determinations of the radius of NGC~2302 have been published by Kharchenko et al. (2005) as $4.8\arcmin$ and by Dias
et al. (2006) as $4.0\arcmin$.  These determinations were part of programs aimed at estimating open cluster parameters on a 
large scale and did not take the artifacts produced by the complex extinction in this field into account, thus leading to larger
radii that accommodate both density peaks. In this dedicated  study we find that the cluster is much more confined than previously thought
with a radius $\sim 2.5{\arcmin}$.

    \begin{figure*}
     \centering
     \includegraphics[width=\columnwidth]{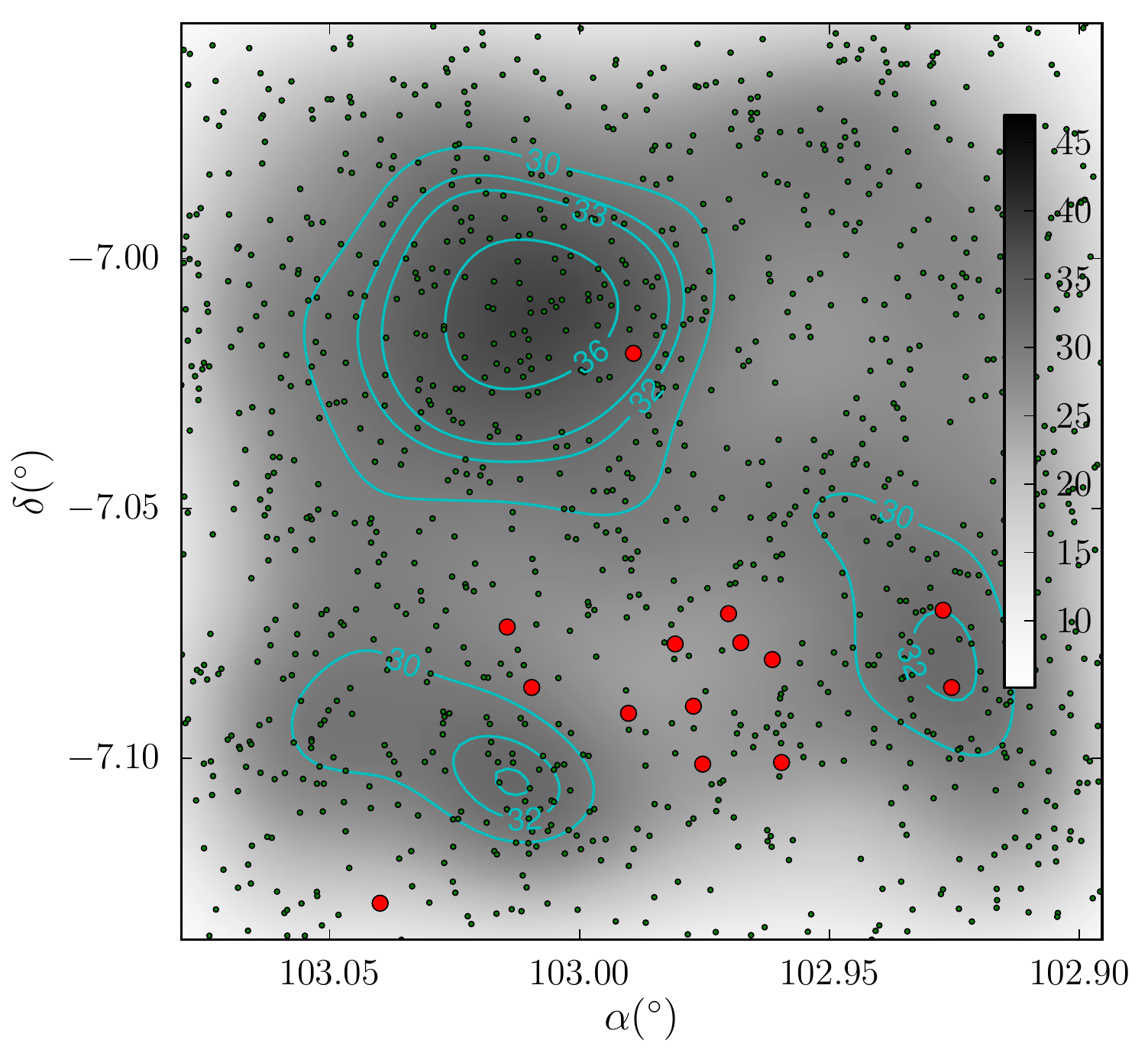}
     \includegraphics[width=\columnwidth]{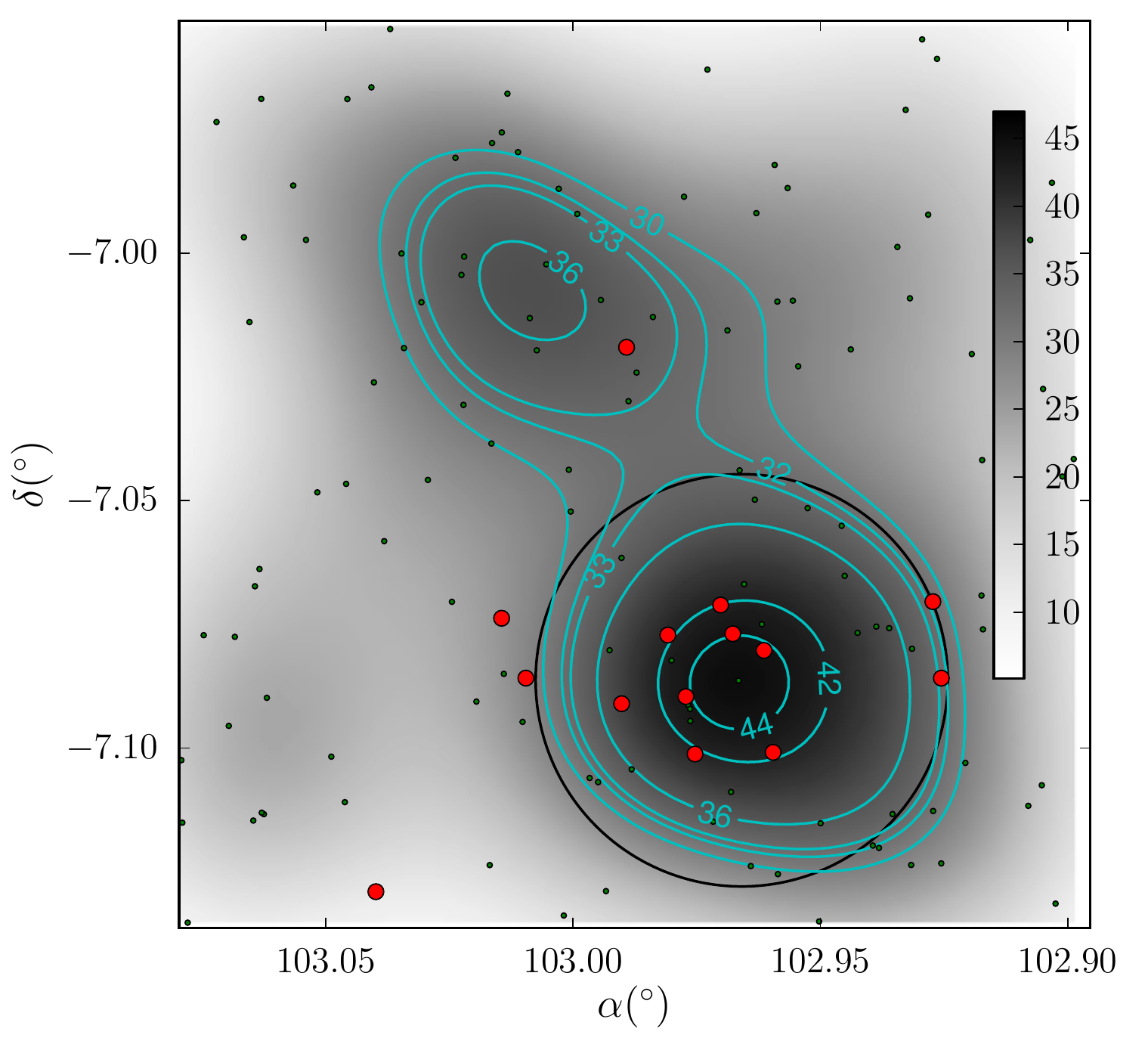}

     \caption{Left panel: Density map constructed using all stars brighter that $V$= 18. Right panel: Density map constructed
              using only stars within 0.1 mag of the cluster's photometric sequence, and brighter than 18 mag.
              The black circle denotes the derived cluster limit. It is centered at  $\alpha_{2000}$ = 06:51:51.820,
              $\delta_{2000}$ = -07:05:10.68 ($\alpha_{2000}$ = 102.965916$^{ \circ}$, $\delta_{2000}$ = -7.086300$^{\circ}$)
              and has a radius of $2.5^{\arcmin}$. In both panels, the brighter and bluer stars ($V <$ 13 and $B-V < 0.6$),
most of which are cluster members, are indicated in red. The density grayscale is in arbitrary units. See text for details.}
     \label{ngc2302dens}
    \end{figure*}

   \subsubsection{Photometric diagrams}

In the subsequent analyses and following the discussion in M01, a standard reddening law with $A_V$ = 3.1 and $E(U-B)/E(B-V)$ = 0.72
has been adopted. Photometric diagrams for several color combinations have been built using our data. They are shown in Figure 3. Although they are
densely contaminated by field stars, a cluster sequence of bright stars ($V \leq$ 18) is easily recognized in all diagrams. The
color-color diagrams (or two-color diagrams, from now on TCD), exhibit a complex structure that shows more than one population, each
affected by a distinct amount of reddening. As previously noted, this is a phenomenon that must be considered when determining
the structural parameters of the cluster, in particular its center and radius.\\ 

Besides the cluster sequence seen at $E(B-V)$ = 0.23, which we discuss later in this section, at higher reddening around
$E(B-V)$ = 0.70, we find a number of early type stars that can also be identified as a blue sequence in the CMDs. These are
{\it \emph{Blue Plume}} stars (Carraro et al. 2005; Moitinho et al 2006), in this case located at approximately $(m-M) = 10.96$ ($d=7.50$ kpc),
tracing the outer arm.\\

    \begin{figure*}
     \centering
     \includegraphics[width=\textwidth]{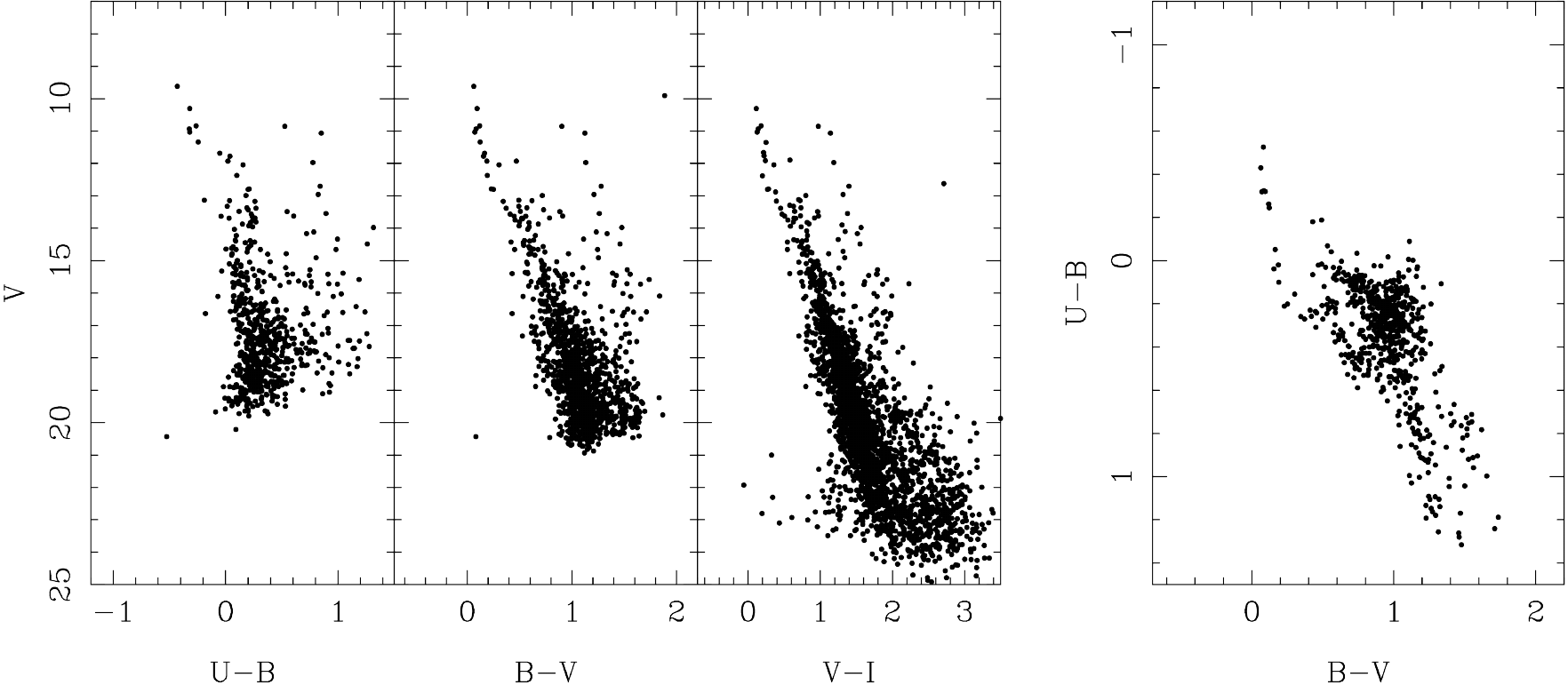}
     \caption{Left panels: CMDs in various color combinations for all stars in the field of NGC~2302 with available $UBVRI$ photometry,
              including the deep $V-I$ vs. $V$ CMD that results from combining the photometry of M01 and the Dupont 2.5m $VI$ data from
              this work. Right panel: $U-B$ vs. $B-V$ color-color diagram for these stars.}
     \label{ngc2302phot}
    \end{figure*}

In Figure 4 we display two TCDs, which are restricted to stars within the cluster radius to alleviate the contamination by field stars.
As a result of the restriction, the cluster sequence is now seen more clearly, allowing a straightforward reddening fit that
yields $E(B-V)$ = 0.23 mag. The fit is illustrated by the superposition of the $\log(t) = 7.90$ (80 Myr) isochrone from Marigo et al. (2008), displaced for a
reddening of $E(B-V)$ = 0.23 mag.\\

    \begin{figure*}
     \centering
     \includegraphics[width=1.5\columnwidth]{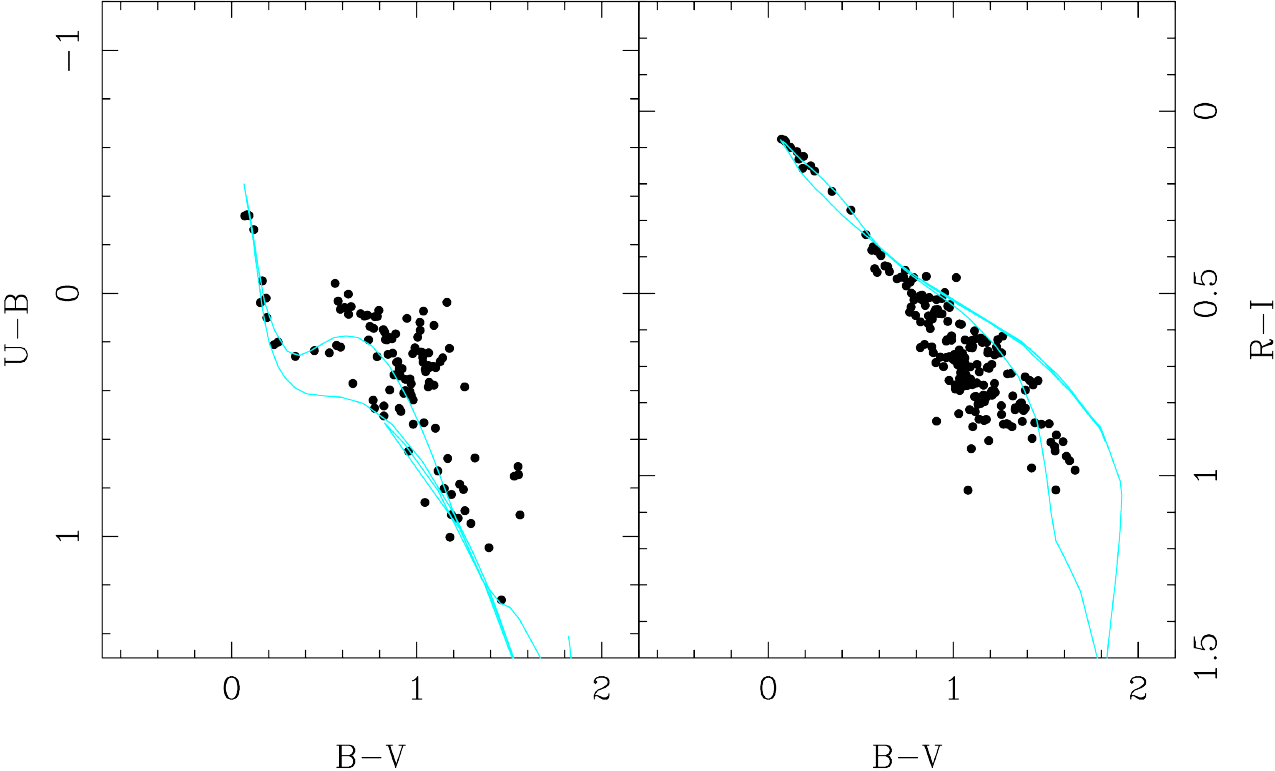}
     \caption{TCD of stars within the adopted radius of NGC~2302. A $\log(t) = 7.90$ (40 Myr) isochrone from Marigo
              et al. (2008), displaced for a reddening of $E(B-V)$ = 0.23, has been superposed.}
     \label{ngc2302tcd}
    \end{figure*}

CMDs for these stars are presented in Figure 5. Again, the cluster sequence is easily identified. Two isochrones from
Marigo et al. (2008), displaced for a reddening of $E(B-V)$ = 0.23, and for a distance modulus of $(m-M)_0 = 10.69$
(distance, $d=1.40$ kpc) have been superposed, along with a $\log(t) = 7.90$ isochrone (80 Myr) and a $\log (t) = 8.00$ isochrone
(100 Myr). We note that only one isochrone was plotted in the TCD to avoid making the figure confusing.\\

    \begin{figure*}
     \centering
     \includegraphics[width=1.3\columnwidth]{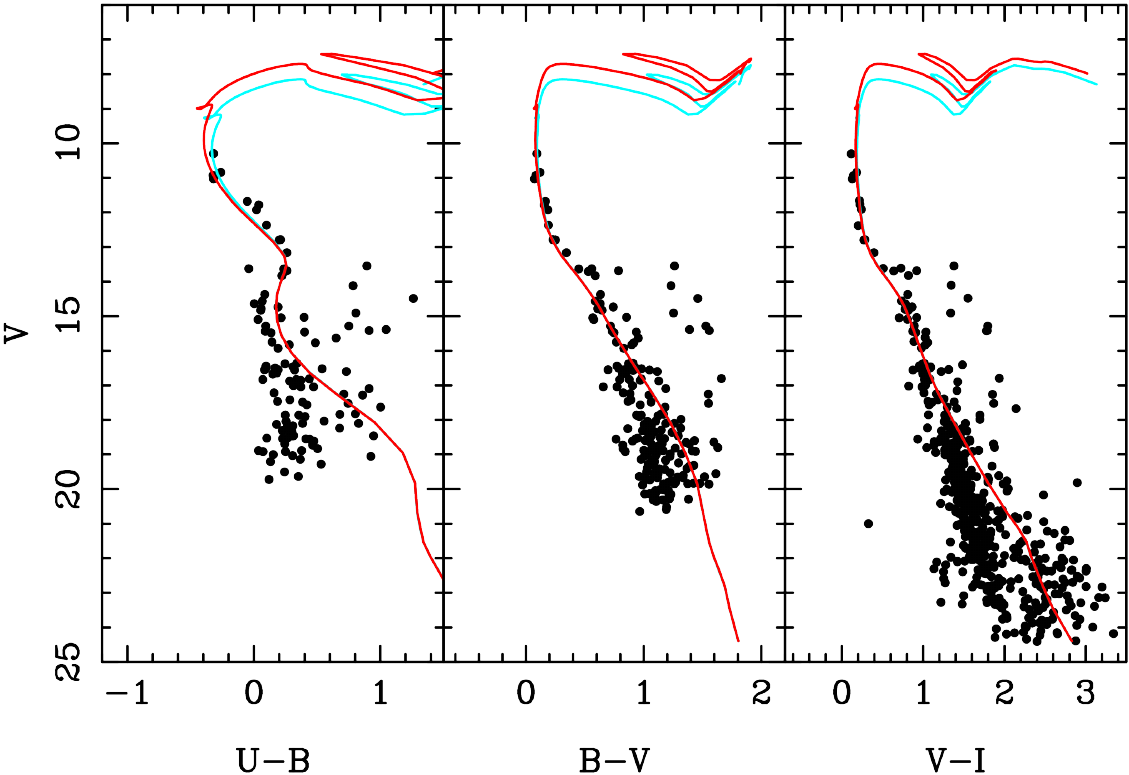}
     \caption{CMDs of stars within the adopted radius of NGC~2302. Two isochrones from Marigo et al. (2008), displaced
              for a reddening of $E(B-V)$ = 0.23 and for a distance modulus of $(m-M)_0 = 10.69$ (distance, $d=1.40$ kpc),
              have been superposed; a $\log(t) = 7.90$ isochrone (80 Myr, red) and a $\log (t)=8.00$ isochrone
              (100 Myr, light blue)}
     \label{ngc2302cmd}
    \end{figure*}

Inspection of the CMDs of Fig.~5 shows that both isochrones, $\log(t) = 7.90-8.00$, provide a similar fit in the middle and righthand panels  (V/B-V and V/V-I diagrams). The V/U-B diagram in the righthand panel indicates that the age cannot be younger than $\log(t) = 7.90$, since this isochrone is already slightly too blue and spans approximately 4 mag in the bright end without stars.  
However, it cannot be much older than $\log(t)= 7.9$, which is favored because for older ages (already at $\log(t)= 8.0$) the isochrone in the TCD fit (right panel) starts missing the earlier type stars.

\subsubsection{Photometric membership analysis}

A preliminary membership analysis was performed based on the photometric data, in order to provide an initial constraint for the 
the determination of the cluster's proper motion. Only stars within a box $8\arcmin \times 8\arcmin$ centered on the cluster
were considered.\\

A two-step approach was followed. First, the unsupervised cluster membership package UPMASK (Krone-Martins \& Moitinho, 2014) was
used. UPMASK can provide cluster membership probabilities in an unsupervised, model-free way, based only on photometric data and
positions. The current version of the package uses the k-means clustering algorithm and requires specifying one single parameter,
$k$, which is the number of stars per k-means cluster. Several trials were performed, and we found that the cluster sequence
was best identified for $k$=31.\\

Given its age, distance, and the fact that it is a poorly populated cluster, it was not expected that UPMASK would yield robust membership
probabilities for NGC~2302 (see the upper panels in Fig.~4 of Krone-Martins \& Moitinho, 2014). Nevertheless, by adopting a value of $k$=31,
l, which is higher than recommended for poorly populated clusters ($k\sim$10), its sequence is identifiable down to $V\sim$ 16 -although at the
price of being very contaminated by field stars.\\

Based on this first overview of cluster memberships, the cluster's CMDs and TCDs were examined, and stars with consistent
positions relative to the cluster's isochrone were selected as probable cluster members. Probable members are depicted in Figure 6, which shows the results of our membership analysis.

    \begin{figure*}
     \centering
      \includegraphics[width=\textwidth]{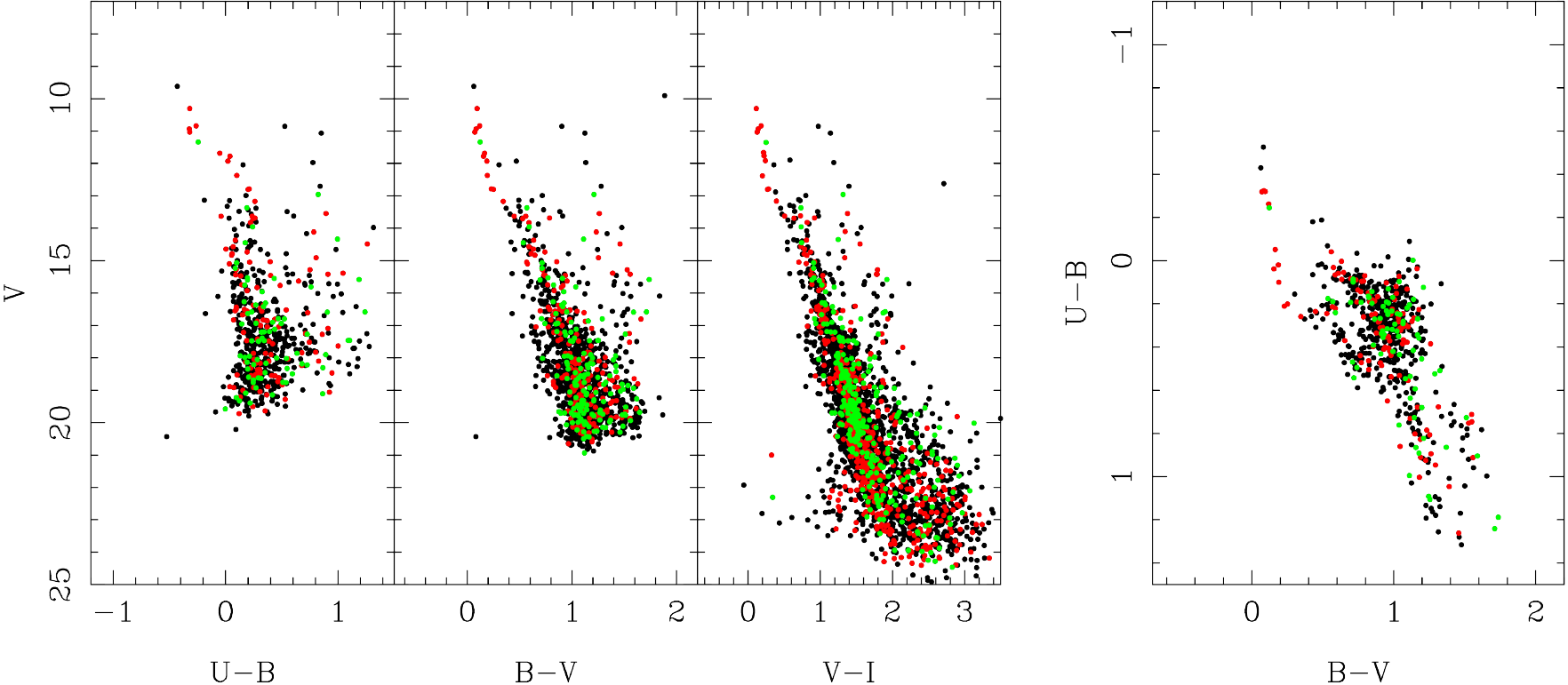}
      \caption{Red: cluster stars; green: secondary peak; black: the rest (includes blue plume)}
    \end{figure*}

\subsection{Radial velocity}

Ninety-nine stars in the field of NGC 2302 were targeted for the RV study. They were selected from the cluster's
upper and mid main sequence in its $V$ {\it vs.} $V-I$ CMD (see figure 3), thus favoring the selection of likely cluster
members for the spectroscopy. They were separated into three brightness groups in order to observe them with the appropriate S/N.
Three Hydra fiber set-ups were created for that purpose, with exposure times of 3, 15, and 23 minutes. Each set-up was 
exposed three times to improve the S/N of the spectra. Seventy-six of the 99 objects were successfully observed, and
their RV extracted as explained in Section 2.3.3. In Table 2 (see Appendix) we present the results from our RV study.
Column 1 is a
running number; Cols. 2 and 3 give the J2000.0 right ascension (RA) and declination (DEC), respectively; Cols.
4 and 5 the J2000.0 RA and DEC, in degrees; Cols. 6 and 7 the radial velocity (RV) and its sigma (RVerr), respectively,
in km s$^{-1}$; Col. 8 the number RV determinations (N), leading to RV and RVerr; and, finally, Col. 9 indicates
whether the star has a PM value in the UCAC4 catalog (see Section 2.2.7).

In  Figure 7 we show the RV histogram for the 76 stars included in Table 2.
Two distinct peaks are seen: a primary peak centered at
$\sim$30 km sec$^{-1}$, and a secondary peak centered at $\sim$60 km sec$^{-1}$. Considering that 11 of the 26 stars
with RV in the neighborhood of the primary peak are indicated as probable cluster members by our photometric membership
analysis, and none is recognized as such in the secondary peak, we interpret the primary peak as the cluster. The mean
RV of these 26 stars turns out to be $+$31.2 km sec$^{-1}$, with a standard error of 0.7 km sec$^{-1}$, a value that we
adopt as the mean RV of NGC~2302.\\

Our result does not compare well with the recent RV determination presented by Dias et al. (2014), who, by
 cross-correlating their membership analysis for NGC 2302 with available RV catalogs, obtained a mean RV for NGC 2302
of $+$43.4$\pm$0.35 km sec$^{-1}$. We think that this discrepancy could be accounted for if we had (incorrectly) included
stars of the secondary RV peak to determine the mean RV.


    \begin{figure*}
    \centering
    \includegraphics[width=12.0cm,angle=0]{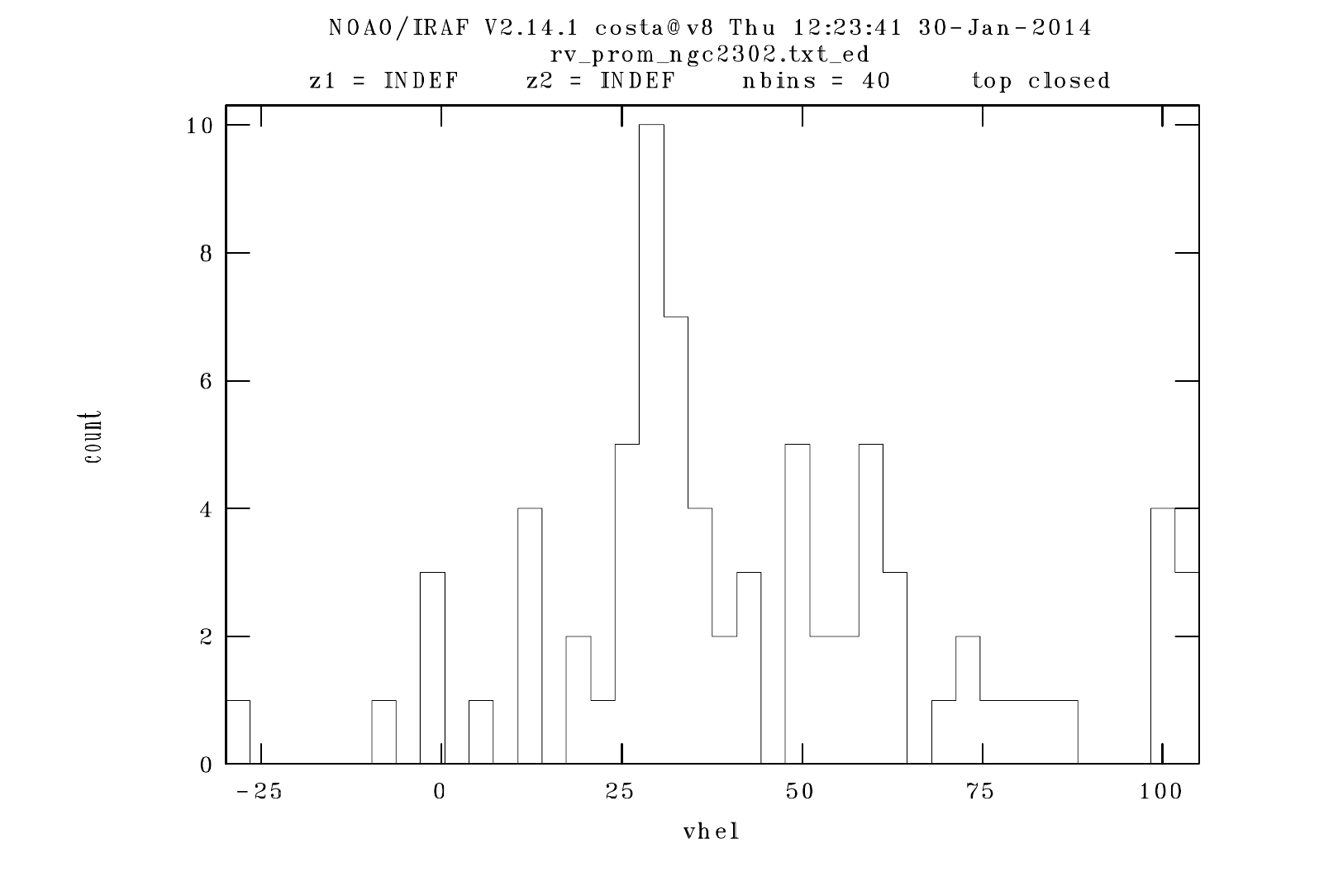}
    \caption{Radial velocity histogram for the 76 NGC 2302 stars observed with HYDRA@CTIO}
    \label{ngc2302RVhistogram}
    \end{figure*}

\subsection{Proper motion}

Four previous determinations of the mean absolute PM for NGC 2302 can be found in the literature:
Glushkova et al. 1997, who used the Four-Million Star Catalog of Positions and Proper Motions;
Lotkin \& Beshenov 2003, who used the Tycho-2 Catalog; Kharchenko et al. 2005, who used the ASCC-2.5 Catalog;
and Dias et al. 2014, who used the UCAC4 Catalog. These results are summarized in Table 3. The results from the {\it \emph{present}} PM survey of the field of NGC~2302 are summarized in Table 4.\\

Through our registration procedure we obtained the mean PM of NGC~2302 relative to the local field of disk stars:\
($\mu_{\alpha} \cos\delta, \mu_{\delta}) = (+0.33,-0.31)$ mas yr$^{-1}$, with standard errors if 0.010 and 0.014 mas yr$^{-1}$,
respectively.          \\

On the other hand, the mean UCAC4 PM of the field stars in the $0.5\degr \times 0.5\degr$ region centered on NGC~2302 is\
($\mu_{\alpha} \cos\delta,\mu_{\delta}) = (-2.42,-1.80)$ mas yr$^{-1}$, with standard errors of 0.41 and 0.40 mas yr$^{-1}$,
respectively. Thus, from this mean field PM and our mean relative PM, we obtain an absolute PM of NGC~2302 of:
($\mu_{\alpha} \cos\delta,\mu_{\delta}) = (-2.09,-2.11)$ mas yr$^{-1}$,  with (adding in quadrature) standard errors of
0.410 and 0.400 mas yr$^{-1}$, respectively.\\

Averaging the UCAC4 PM of the 31 probable cluster members, selected from the peak of the RV histogram of NGC~2302
and supplemented by high-probability cluster members (with no RV determination) selected by means of a photometric
membership analysis, the absolute PM of NGC~2302 can be obtained directly from that catalog, and is\
($\mu_{\alpha} \cos\delta,\mu_{\delta}) = (-2.08,-2.37)$ mas yr$^{-1}$, with standard errors of 0.72 and 0.62
mas yr$^{-1}$, respectively. 
In Figure 8 we show the PM diagram for stars with bona fide RV {\it and} PM. We note that, for
clarity, non-members with very large PMs were not included in the figure.


    \begin{figure*}
    \centering
    \includegraphics[width=12.0cm,angle=0]{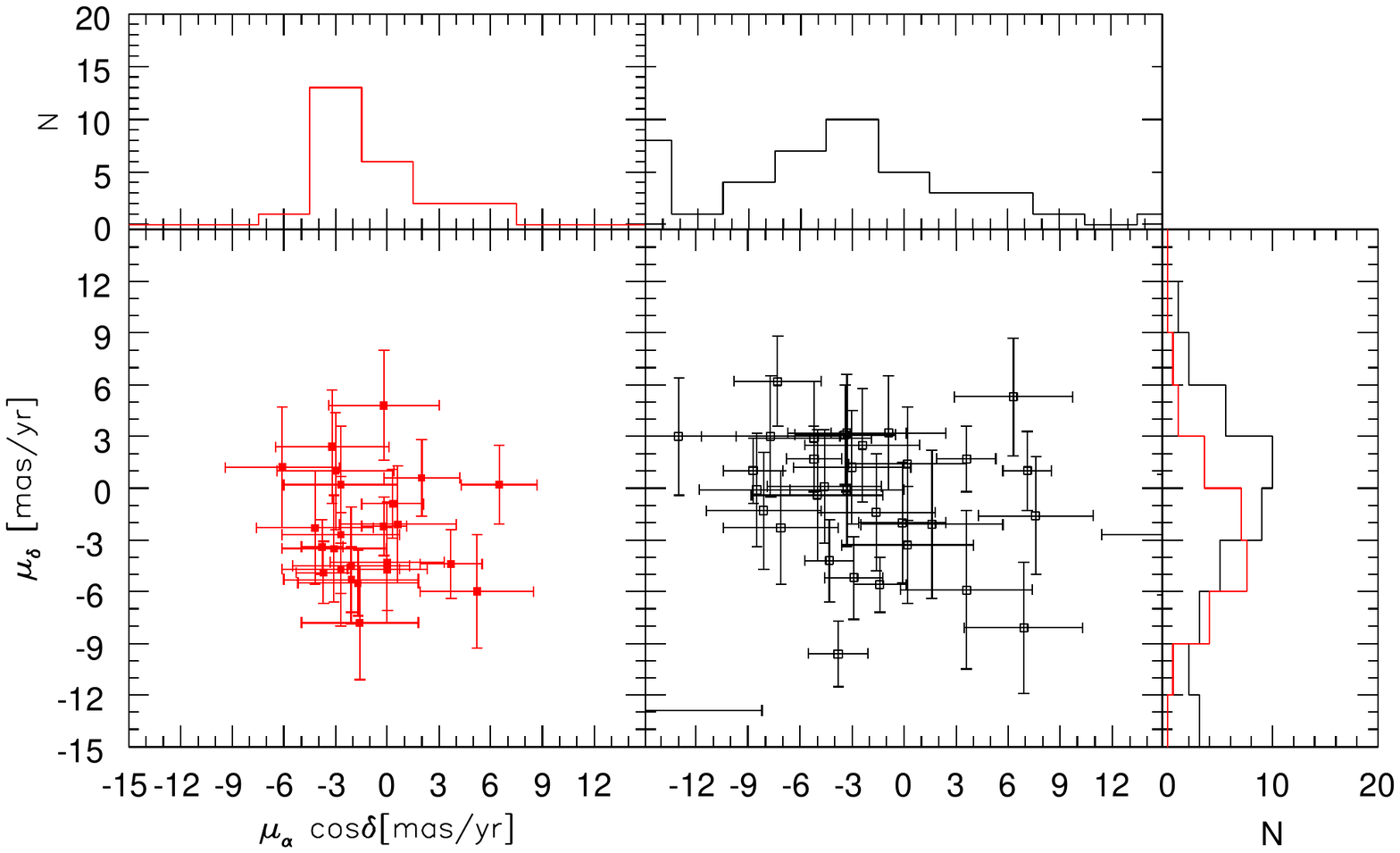}
    \caption{Proper motion diagram for stars with bona-fide RV {\it \emph{and}} PM. Red dots depict stars that we
             considered as cluster members for all purposes, and black dots those considered as non-members.
             We note that, for clsrity, non-members with very large PMs were not included in the figure.}
    \label{ngc2302PMdiagram}
    \end{figure*}

\begin{table*}
\begin{flushleft}
\caption{Proper motion results for NGC~2302 from this work}
\begin{tabular}{lcccc}
\hline
\noalign{\smallskip}
{Source} & $\mu_{\alpha}$cos$\delta$ & Std.err.     & $\mu_{\delta}$ & Std.err. \\
         &mas yr$^{-1}$              &mas yr$^{-1}$ &mas yr$^{-1}$   &mas yr$^{-1}$ \\
\noalign{\smallskip}
\hline
\noalign{\smallskip}
Mean absolute PM for NGC~2302 directly from UCAC4                                           & $-$2.08 &0.72 & $-$2.37 &0.62  \\
PM of NGC~2302 relative to the local field of UCAC4 stars                                   & $+$0.33 &0.010& $-$0.31 &0.014 \\
Absolute PM for NGC~2302 (our procedure)                                                    & $-$2.09 &0.41 & $-$2.11 &0.40 \\
\noalign{\smallskip}
\hline
\end{tabular}
\end{flushleft}
\end{table*}

\section{Orbital parameters}

The cluster's kinematic data (RV and absolute PM) and distance derived in this work allowed us to determine the space
motion of NGC~2302  and its orbital parameters. These, in turn, allowed us to check the kinematic data and the cluster age, since we expect that such a young cluster close to the Sun moves in an almost spherical orbit.  
As a first step, we translated the RV and PM components into Galactic Cartesian velocity. To do so,
we adopted the heliocentric distance derived in this work for NGC~2302, allowing for an uncertainty of $\sim 15\%$.
The results are presented in Table 5, where for three different distances, we list the Cartesian velocity components
U, V, and W, and their values once shifted to the local standard of rest and corrected for solar motion.
The procedure and adopted parameters are the same as in Bedin et al. (2006), where we derived the orbit of the old open
cluster NGC~6791.

\begin{table}
\caption{Input conditions for orbit calculation. Distances are in kpc, and velocities in km s$^{-1}$. }
\centering
\label{orbIN}
\begin{tabular}{cccccc}
\hline
\noalign{\smallskip}
d   & U         & V           & W  ($=$ Z)  & $\Pi$     & $\Theta$    \\
\hline
1.2 & $-32\pm 4$ & $+1.0\pm 3 $ & $-20\pm 1$ & $-28\pm 4$ & $222\pm 4 $ \\
1.4 & $-35\pm 4$ & $+0.6\pm 2 $ & $-21\pm 2$ & $-30\pm 4$ & $221\pm 3 $ \\
1.6 & $-37\pm 5$ & $-0.5\pm 2 $ & $-20\pm 2$ & $-32\pm 4$ & $220\pm 3 $ \\
\noalign{\smallskip}
\hline
\end{tabular}
\end{table}

To integrate the orbit of NGC~2302, we adopted the model of
Allen \& Santillan (1991) as model for the MW gravitational
potential. This potential is  time-independent, axisymmetric, fully  analytic, and mathematically very
simple. It was constructed to fit a certain Galactic rotation curve (it assumes densities for the bulge, disk, and halo
whose combined gravitational force fits a rotation curve consistent with observations); and given Galactocentric distance and
rotation velocity for the Sun.  It is reasonable to believe that the Galactic
potential does not change much during the lifetime of a young cluster (less than 100 Myr in the case of NGC~2302), so
that the derived parameters for its orbit, such as the apo- and perigalacticon, can be considered to be good  estimates.
This potential has already been used successfully to  derive the Galactic orbits of nearby stars  (Bensby et al. 2014), open
clusters (Carraro \& Chiosi 1994, Carraro et al. 2006), and also disk and halo globular clusters (Odenkirchen \& Brosche 1992,
Milone et al. 2006).\\

The orbit-integration routine used was a fifteenth-order symmetric, simplectic Runge-Kutta, using the Radau scheme (Everhart 1985).
This guarantees conservation of energy and momentum at a level of $10^{-12}$ and $10^{-9}$, respectively, over the whole orbit
integration. The orbits calculated were integrated back in time for 0.5 Gyr and  are shown in Figure 9.\\

The resulting orbital parameters are summarized in Table 6 where Col. (1) lists the adopted cluster's
heliocentric distance, Cols. (2) and (3) the apo- and peri-center of the orbit, Col. (4) the maximum vertical distance
reached, and  Col. (5) the eccentricity, defined as $(R_a-R_p)/(R_a+R_p)$. The resulting orbits are shown in Figure 8
for each adopted value of the cluster distance. The lefthand panels show the orbit in the plane of the Galactic disk, while
the righthand ones show the orbits in its meridional plane.\\

\begin{table}
\caption{Orbit parameters for the three different distances.
Units:\
$d${[kpc]},
$L_{\rm z}${[kpc km s$^{-1}$]},
$E_{\rm tot}${[${\rm 10 \times km^2 s^{-2}}$]},
$P ${[Myr]},
$R_{\rm a}${[kpc]},
$R_{\rm p}${[kpc]},
$z_{\rm max}${[kpc]},
$e${[pure number]}.
}
\centering
\label{orbOUT}
\begin{tabular}{ccccc}
\hline
     $d$ & $R_{\rm a}$& $R_{\rm p}$  &$z_{\rm max}$&  $e$   \\
\hline
1.2 &  10.20 & 9.30 & 0.31  & 0.05 \\
1.4 &  10.32 & 9.46 & 0.31  & 0.04 \\
1.6 &  10.44 & 9.62 & 0.32  & 0.04 \\
\hline                                   
\end{tabular}
\end{table}

  \begin{figure}
    \centering
\includegraphics[width=\columnwidth]{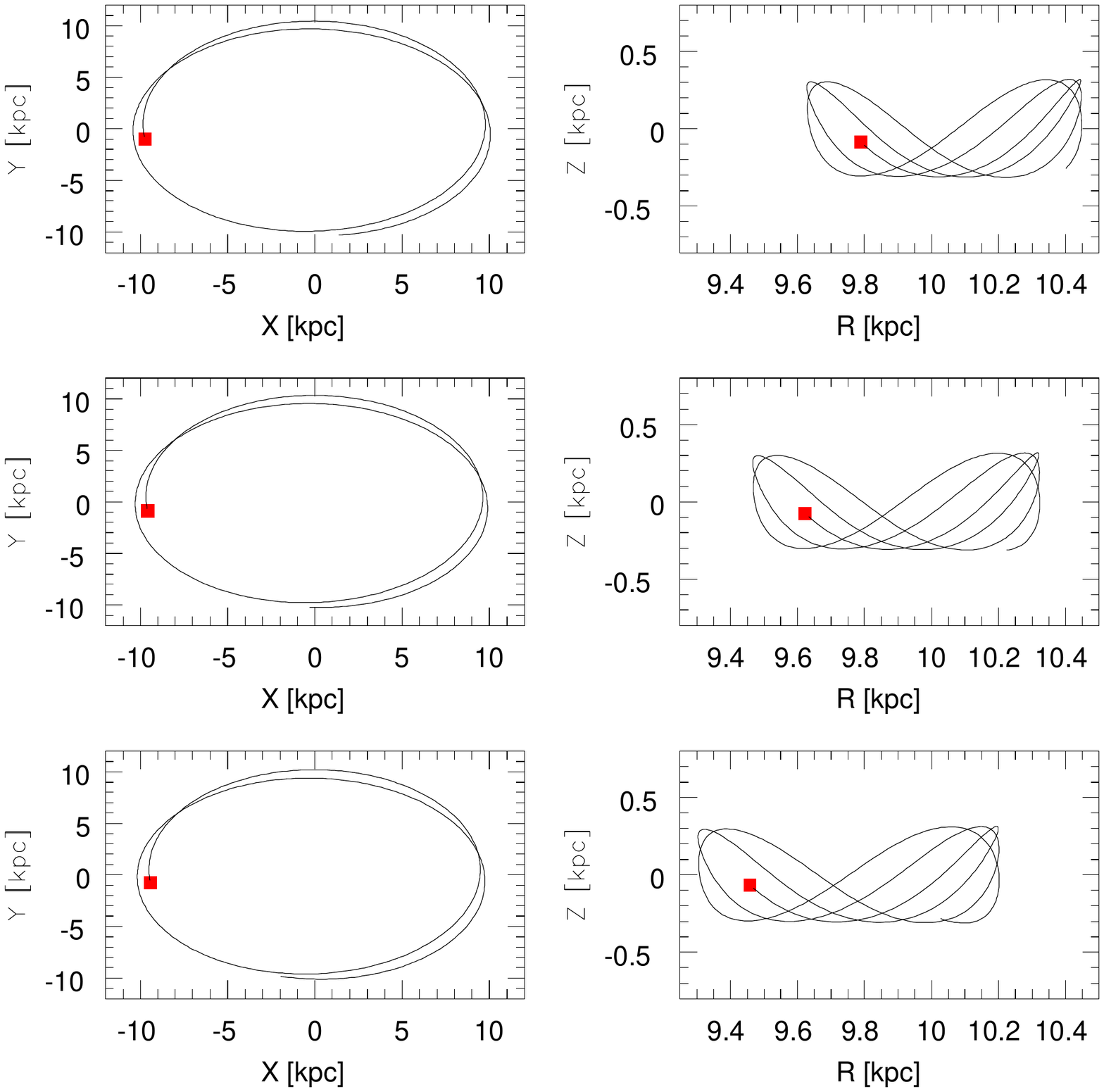}
    \caption{Orbits calculated back in time for 0.5 Gyr, for assumed cluster distances of 1.2, 1.4, and 1.6 kpc. The red squares indicate the current location of the cluster.}
    \label{orbit}
  \end{figure}
    
\noindent
Inspecting the shape of the orbits and the resulting orbital parameters, we conclude that NGC~2302 is a typical Population I object.
The eccentricity of the orbit is low, as is its epicyclical amplitude. The absolute maximum distance from the plane is that expected for a  Population I object as well.

\section{Discussion and conclusions}

In this work we have presented a new photometric, astrometric, and spectroscopic data set for the young open cluster NGC2302,
located in the third Galactic quadrant, where we aim to map the spiral structure. Because this paper is the first in a series on the
LOA, it was mostly devoted to describing the aim of the program, data acquisition, and reduction, and it told how we extracted the kinematic
data (PM and RV). In forthcoming papers, we shall present data for more young clusters of this Galactic sector and also
discuss the consequences of our findings for the spiral structure of the MW.\\

\noindent
Although results for NGC 2302 have been included in a number of large scale studies of open cluster surveys with a
variety of purposes, we stress that this is the first deep and comprehensive study of NGC 2302, including spectroscopy for
a large number of stars, and second-epoch photometry with a baseline of more than 12 years. This allowed us to derive
robust and updated estimates of the fundamental parameters of NGC~2302: age, reddening, distance, mean radial velocity, and
mean absolute PM. By means of our kinematical data, we could derive the cluster's orbital parameters, and confirm
that it is a classical Population I object.\\

NGC~2302 is 1.4 kpc away from the Sun. At this distance, its position in the third Galactic quadrant can be compatible both 
with the Local and the Perseus arm.  In V\'azquez et al. (2008), favored a scenario in which the Perseus arm is broken
in the third quadrant, and most of the young
population in the sector is probably associated with the Local Arm, which extends in the third quadrant up to the distant
outer (Cygnus) arm. This scenario, however, was entirely based on photometric data and on the spatial distribution of the tracers
(star clusters) in the plane of the Galaxy. The present program will add the much needed kinematical information for our sample of
star clusters, and it allowed us to derive the bulk motion of these tracers (which we already found can spatially define a spiral feature).

\begin{acknowledgements}
EC acknowledges support by the Fondo Nacional de Investigaci\'on Cient\'ifica y Tecnol\'ogica (proyecto No. 1110100 Fondecyt)
and the Chilean Centro de Excelencia en Astrof\'isica y Tecnolog\'ias Afines (PFB 06). The  useful comments of the anonymous referee are deeply acknowledged.
\end{acknowledgements}


\begin{appendix} 
\setcounter{table}{1}
\begin{table*}
        \tabcolsep 0.10truecm
        \caption{Radial velocities. See text for details}
        \begin{tabular}{lcccccccc}
        \hline
        \noalign{\smallskip}
No & RA (J2000.0) & DEC (J2000.0) & RA (J2000.0) & DEC (J2000.0) & RV & RVerr & N & PM\\
   & h m s & $\degr$$\arcmin$$\arcsec$ & $\degr$ & $\degr$ & km s$^{-1}$ & km s$^{-1}$ &   &   \\
        \noalign{\smallskip}
        \hline
        \noalign{\smallskip}
1 & 6:51:35.08 & $-$6:59:34.90 & 102.896  & $-$6.993 & 30.24  & 2.98 & 18 & UCAC4 \\
2 & 6:51:35.08 & $-$7:06:29.00 & 102.896  & $-$7.108 & 40.45  & 1.67 & 22 & UCAC4 \\
3 & 6:51:35.71 & $-$7:02:29.60 & 102.899  & $-$7.042 & 82.44  & 3.13 & 18 &       \\
4 & 6:51:37.20 & $-$7:01:38.70 & 102.905  & $-$7.023 & 100.70 & 3.43 & 18 &       \\
5 & 6:51:37.70 & $-$6:59:36.40 & 102.907  & $-$6.993 & 47.70  & 1.02 & 20 & UCAC4 \\
6 & 6:51:40.15 & $-$7:02:30.40 & 102.917  & $-$7.042 & 53.37  & 2.33 & 21 & UCAC4 \\
7 & 6:51:40.18 & $-$7:04:09.00 & 102.917  & $-$7.069 & 37.04  & 3.01 & 20 & UCAC4 \\
8 & 6:51:40.29 & $-$7:03:12.80 & 102.919  & $-$7.054 & 60.11  & 1.39 & 21 & UCAC4 \\
9 & 6:51:40.32 & $-$6:57:45.40 & 102.918  & $-$6.963 & 63.24  & 1.57 & 19 & UCAC4 \\
10& 6:51:40.66 & $-$7:01:13.20 & 102.919  & $-$7.020 & 58.23  & 1.91 & 21 & UCAC4 \\
11& 6:51:40.97 & $-$7:06:10.80 & 102.921  & $-$7.103 & 33.00  & 1.60 & 19 & UCAC4 \\
12& 6:51:41.67 & $-$7:05:38.40 & 102.924  & $-$7.094 & 55.64  & 4.51 & 13 & UCAC4 \\
13& 6:51:42.13 & $-$7:05:09.10 & 102.926  & $-$7.086 & 32.27  & 3.01 & 17 & UCAC4 \\
14& 6:51:42.54 & $-$7:04:13.40 & 102.927  & $-$7.070 & 35.77  & 2.51 & 17 & UCAC4 \\
15& 6:51:44.27 & $-$6:59:55.30 & 102.934  & $-$6.999 & 30.15  & 1.89 & 23 & UCAC4 \\
16& 6:51:45.13 & $-$7:00:47.90 & 102.938  & $-$7.013 & 104.25 & 2.06 & 17 &       \\
17& 6:51:45.16 & $-$7:07:12.60 & 102.938  & $-$7.120 & 33.16  & 3.88 & 16 & UCAC4 \\
18& 6:51:45.22 & $-$6:59:28.90 & 102.938  & $-$6.991 & 60.46  & 3.20 & 17 & UCAC4 \\
19& 6:51:46.20 & $-$7:04:36.20 & 102.943  & $-$7.077 & 30.57  & 1.91 & 18 & UCAC4 \\
20& 6:51:46.82 & $-$7:03:54.70 & 102.945  & $-$7.065 & 30.67  & 1.15 & 19 & UCAC4 \\
21& 6:51:47.26 & $-$7:05:05.60 & 102.947  & $-$7.085 & 85.53  & 3.04 & 20 &       \\
22& 6:51:47.51 & $-$7:01:23.20 & 102.948  & $-$7.023 & 45.44  & 1.15 & 22 & UCAC4 \\
23& 6:51:47.99 & $-$7:06:54.70 & 102.999  & $-$7.115 & 74.26  & 3.56 & 18 &       \\
24& 6:51:49.34 & $-$7:00:34.40 & 102.956  & $-$7.009 & 37.68  & 1.21 & 18 & UCAC4 \\
25& 6:51:49.65 & $-$7:07:16.50 & 102.957  & $-$7.121 & 29.87  & 1.65 & 21 & UCAC4 \\
26& 6:51:50.07 & $-$6:58:57.40 & 102.959  & $-$6.983 & 56.95  & 1.99 & 22 & UCAC4 \\
27& 6:51:50.30 & $-$7:06:03.10 & 102.959  & $-$7.101 & 18.04  & 2.52 & 16 & UCAC4 \\
28& 6:51:50.74 & $-$7:04:48.90 & 102.961  & $-$7.080 & 34.87  & 1.98 & 17 & UCAC4 \\
29& 6:51:51.38 & $-$7:07:25.80 & 102.964  & $-$7.124 & 36.55  & 1.46 & 17 & UCAC4 \\
30& 6:51:52.23 & $-$7:03:29.40 & 102.968  & $-$7.058 & 43.49  & 1.48 & 20 & UCAC4 \\
31& 6:51:52.51 & $-$7:00:56.10 & 102.969  & $-$7.016 & 25.20  & 1.58 & 21 & UCAC4 \\
32& 6:51:52.84 & $-$7:04:15.90 & 102.970  & $-$7.071 & 23.16  & 1.91 & 16 & UCAC4 \\
33& 6:51:53.21 & $-$7:06:53.60 & 102.972  & $-$7.115 & 51.97  & 3.21 & 18 & UCAC4 \\
34& 6:51:54.08 & $-$7:06:04.30 & 102.975  & $-$7.101 & 26.43  & 2.46 & 18 & UCAC4 \\
35& 6:51:54.41 & $-$7:05:28.50 & 102.977  & $-$7.091 & 25.09  & 1.96 & 18 & UCAC4 \\
36& 6:51:54.53 & $-$7:05:22.50 & 102.977  & $-$7.089 & 42.81  & 2.24 & 16 & UCAC4 \\
37& 6:51:55.10 & $-$7:03:27.50 & 102.979  & $-$7.058 & -1.05  & 1.42 & 20 &       \\
38& 6:51:55.13 & $-$7:04:28.60 & 102.979  & $-$7.075 & 11.08  & 1.63 & 22 & UCAC4 \\
39& 6:51:55.40 & $-$7:04:37.70 & 102.981  & $-$7.077 & 20.42  & 1.75 & 14 & UCAC4 \\
40& 6:51:56.58 & $-$7:02:24.10 & 102.986  & $-$7.040 & 12.73  & 2.79 & 21 & UCAC4 \\
41& 6:51:56.61 & $-$7:01:38.70 & 102.986  & $-$7.027 & -8.34  & 1.22 & 20 &       \\
42& 6:51:57.15 & $-$6:57:20.10 & 102.988  & $-$6.956 & 76.49  & 1.71 & 15 &       \\
43& 6:51:57.16 & $-$7:06:15.40 & 102.988  & $-$7.104 & 28.51  & 1.50 & 17 & UCAC4 \\
44& 6:51:57.30 & $-$7:04:01.30 & 102.989  & $-$7.067 & 101.75 & 1.62 & 21 &       \\
45& 6:51:57.41 & $-$7:01:08.30 & 102.989  & $-$7.019 & 64.15  & 3.83 & 15 & UCAC4 \\
46& 6:51:57.65 & $-$7:05:27.70 & 102.990  & $-$7.091 & 24.14  & 2.47 & 14 & UCAC4 \\
47& 6:51:58.18 & $-$6:58:05.50 & 102.992  & $-$6.968 & 28.45  & 1.98 & 22 & UCAC4 \\
48& 6:51:58.41 & $-$7:07:44.10 & 102.993  & $-$7.129 & 30.03  & 1.88 & 22 & UCAC4 \\
49& 6:51:58.79 & $-$7:06:24.80 & 102.995  & $-$7.107 & 31.60  & 1.66 & 17 & UCAC4 \\
50& 6:51:59.20 & $-$7:06:21.70 & 102.997  & $-$7.106 & 31.75  & 1.53 & 20 & UCAC4 \\
51& 6:52:00.25 & $-$7:03:37.20 & 103.001  & $-$7.060 & 32.51  & 1.34 & 23 & UCAC4 \\
52& 6:52:00.76 & $-$7:00:59.70 & 103.003  & $-$7.017 & 60.16  & 1.37 & 22 & UCAC4 \\
53& 6:52:03.19 & $-$6:58:03.80 & 103.013  & $-$6.968 & 13.38  & 1.91 & 18 & UCAC4 \\
54& 6:52:03.47 & $-$7:04:25.50 & 103.014  & $-$7.074 & 13.25  & 1.16 & 19 & UCAC4 \\
55& 6:52:03.97 & $-$7:02:18.40 & 103.017  & $-$7.038 & 41.23  & 1.92 & 17 & UCAC4 \\
56& 6:52:04.05 & $-$7:07:25.10 & 103.017  & $-$7.124 & 30.30  & 3.26 & 14 & UCAC4 \\
57& 6:52:04.33 & $-$7:04:57.40 & 103.018  & $-$7.083 & -2.70  & 1.26 & 19 &       \\
58& 6:52:05.29 & $-$7:00:02.30 & 103.022  & $-$7.001 & 25.70  & 7.19 & 7  & UCAC4 \\
59& 6:52:05.95 & $-$7:01:04.00 & 103.025  & $-$7.018 & -29.95 & 1.58 & 18 &       \\
60& 6:52:07.05 & $-$7:02:44.90 & 103.029  & $-$7.046 & 100.56 & 3.74 & 18 &       \\
 \noalign{\smallskip}
 \hline
 \end{tabular}
 \end{table*}

\addtocounter{table}{-1}
\begin{table*}
        \tabcolsep 0.10truecm
        \caption{Radial velocities. See text for details (continued)}
        \begin{tabular}{lcccccccc}
        \hline
        \noalign{\smallskip}
No & RA (J2000.0) & DEC (J2000.0) & RA (J2000.0) & DEC (J2000.0) & RV & RVerr & N & PM\\
   & h m s & $\degr$$\arcmin$$\arcsec$ & $\degr$ & $\degr$ & km s$^{-1}$ & km s$^{-1}$ &   &   \\
        \noalign{\smallskip}
        \hline
        \noalign{\smallskip}
61& 6:52:08.77 & $-$7:03:53.00 & 103.037  & $-$7.065 & 71.83  & 4.40 & 17 &       \\
62& 6:52:09.67 & $-$7:01:33.90 & 103.040  & $-$7.026 & -2.84  & 1.43 & 20 &       \\
63& 6:52:09.79 & $-$6:57:59.20 & 103.041  & $-$6.966 & 61.34  & 3.61 & 19 & UCAC4 \\
64& 6:52:11.99 & $-$7:04:47.90 & 103.049  & $-$7.080 & 99.32  & 2.14 & 18 &       \\
65& 6:52:12.93 & $-$7:07:49.10 & 103.054  & $-$7.130 & 50.62  & 3.27 & 19 & UCAC4 \\
66& 6:52:13.55 & $-$7:07:23.50 & 103.056  & $-$7.123 & 33.39  & 1.58 & 22 & UCAC4 \\
67& 6:52:13.58 & $-$6:59:10.70 & 103.056  & $-$6.986 & 60.41  & 3.59 & 18 & UCAC4 \\
68& 6:52:14.34 & $-$7:05:54.30 & 103.059  & $-$7.098 & 6.89   & 1.16 & 21 &       \\
69& 6:52:15.01 & $-$7:06:48.10 & 103.063  & $-$7.113 & 30.26  & 2.59 & 20 & UCAC4 \\
70& 6:52:15.13 & $-$6:58:07.60 & 103.063  & $-$6.969 & 105.06 & 7.12 & 19 &       \\
71& 6:52:17.11 & $-$7:03:56.80 & 103.071  & $-$7.066 & 59.27  & 1.56 & 18 & UCAC4 \\
72& 6:52:18.70 & $-$7:08:07.00 & 103.078  & $-$7.135 & 48.47  & 3.70 & 15 & UCAC4 \\
73& 6:52:18.88 & $-$7:02:43.50 & 103.079  & $-$7.045 & 48.49  & 2.93 & 19 &       \\
74& 6:52:18.97 & $-$7:06:54.30 & 103.079  & $-$7.115 & 81.10  & 4.52 & 9  &       \\
 \noalign{\smallskip}
 \hline
 \end{tabular}
 \end{table*}

\end{appendix}


\begin{thebibliography}{}
\bibitem[Allen \& Santillan (1991)]{all91} Allen, C., \& Santillan, A.\ 1991, \rmxaa, 22, 255
\bibitem[Arias et al. (1995)]{ari95} Arias, E.~F., Charlot, P., Feissel, M., \& Lestrade, J.-F.\ 1995, \aap, 303, 604
\bibitem[Avesova(1985)]{ave85} Avedisova, V.~S.\ 1985, Pisma v Astronomicheskii Zhurnal, 11, 448
\bibitem[Bedin et al. (2006)]{bed06} Bedin, L.~R., Piotto, G., Carraro, G., King, I.~R.,
                                 \& Anderson, J.\ 2006, \aap, 460, L27
\bibitem[Bensby et al. (2014)]{ben14} Bensby, T., Feltzing, S., \& Oey, M.~S.\ 2014, \aap, 562, 71
\bibitem[Brand \& Blitz (1993)]{bra93} Brand, J., \& Blitz, L.\ 1993, \aap, 275, 67
\bibitem[Burton (1985)]{bur85} Burton, W.B.\ 1985, \aaps, 62, 365
\bibitem[Carraro et al. (2005)]{car05} Carraro, G., V{\'a}zquez, R.~A., Moitinho, A., \& Baume, G.\ 2005, \apjl, 630, L153
\bibitem[Carraro et al. (2006)]{car06} Carraro, G., Villanova, S., Demarque, P., et al.\ 2006, \apj, 643, 1151
\bibitem[Carraro \& Chiosi (1994)]{car94} Carraro, G., \& Chiosi, C.\ 1994, \aap, 288, 751
\bibitem[Costa et al. (2005)]{cos05} Costa, E., M{\'e}ndez, R.~A., Jao, W.-C., et al.\ 2005, \aj, 130, 337
\bibitem[Costa et al. (2006)]{cos06} Costa, E., M{\'e}ndez, R.~A., Jao, W.-C., et al.\ 2006, \aj, 132, 1234
\bibitem[Costa et al. (2009)]{cos09} Costa, E., M{\'e}ndez, R.~A., Pedreros, M.~H., et al.\ 2009, \aj, 137, 4339
\bibitem[Costa et al. (2011)]{cos11} Costa, E., M{\'e}ndez, R.~A., Pedreros, M.~H., et al.\ 2011, \aj, 141, 136
\bibitem[(de la Fuente Marcos \& de la Fuente Marcos (2009))]{deL09} de la Fuente Marcos, R., \& de la Fuente Marcos, C.\ 2009, \aap, 500, L13
\bibitem[Dias et al. (2005)]{dia05} Dias, W.~S., \& L{\'e}pine, J.~R.~D.\ 2005, \aap, 629, 825
\bibitem[Dias et al. (2006)]{dia06} Dias, W.~S., Assafin, M., Fl{\'o}rio, V., Alessi, B.~S., \& L{\'{\i}}bero,
                                 V.\ 2006, \aap, 446, 949
\bibitem[Dias et al. (2014)]{dia14} Dias, W.~S., Monteiro, H., Caetano, T.~C., et al.\ 2014, \aap, 564, AA79
\bibitem[Everhart (1985)]{eve85} Everhart, E.\ 1985, Dynamics of Comets: Their Origin and Evolution, Proceedings of   IAU Colloq.~83, held in Rome, Italy, June 11-15, 1984.~ Edited by Andrea Carusi and Giovanni B.~Valsecchi.~Dordrecht: Reidel, Astrophysics and Space Science Library.~Volume 115, 1985, p.185, 185
\bibitem[Glushkova et al. (1997)]{glu97} Glushkova, E.~V., Zabolotskikh, M.~V., Rastorguev, A.~S., Uglova, I.~M.,   \& Fedorova, A.~A.\ 1997, Astronomy Letters, 23, 71
\bibitem[Janes \& Adler (1982)]{jan82} Janes, K., \& Adler, D.\ 1982, \apjs, 49, 425
\bibitem[Jao (2009)]{jao09} Jao, W.-C., 2009, Private communication to Edgardo Costa
\bibitem[Kharchenko et al. (2005)]{kar05} Kharchenko, N.~V., Piskunov, A.~E., R{\"o}ser, S., Schilbach, E.,
                                 \& Scholz, R.-D.\ 2005, \aap, 438, 1163
\bibitem[Kharchenko et al. (2009)]{kar09a} Kharchenko, N.~V., Berczik, P., Petrov, M.~I., et al.\ 2009, \aap, 495, 807
\bibitem[Kharchenko et al. (2009)]{kar09b} Kharchenko, N.~V., Piskunov, A.~E., R{\"o}ser, S., et al.\ 2009, \aap, 504, 681
\bibitem[Krone-Martins \& Moitinho (2014)]{kro14} Krone-Martins, A., \& Moitinho, A.\ 2014, \aap, 561, AA57
\bibitem[Loktin \& Beshenov (2003)]{lot03} Loktin, A.~V., \& Beshenov, G.~V.\ 2003, Astronomy Reports, 47, 6
\bibitem[Marigo et al. (2008)]{mar08} Marigo, P., Girardi, L., Bressan, A., et al.\ 2008, \aap, 482, 883
\bibitem[May et al. (1988)]{may88} May, J., Murphy, D.~C., \& Thaddeus, P.\ 1988, \aap, 73, 51
\bibitem[Mendez \& van Altena (1996)]{men96} Mendez, R.~A., \& van Altena, W.~F.\ 1996, \aap, 112, 655
\bibitem[Milone et al. (2006)]{mil06} Milone, A.~P., Villanova, S., Bedin, L.~R., et al.\ 2006, \aap, 456, 517
\bibitem[Moffat \& Vogt (1975)]{mof75} Moffat, A.~F.~J., \& Vogt, N.\ 1975, \aap, 20, 85
\bibitem[Moffat et al. (1979)]{mof79} Moffat, A.~F.~J., Jackson, P.~D., \& Fitzgerald, M.~P.\ 1979, \aap, 38, 197
\bibitem[Moitinho et al. (1997)]{moi97} Moitinho, A., Alfaro, E.~J., Yun, J.~L., \& Phelps, R.~L.\ 1997, \aap, 113, 1359
\bibitem[Moitinho (2001)]{moi01} Moitinho, A.\ 2001, \aap, 370, 436
\bibitem[Moitinho et al. (2006)]{moi06} Moitinho, A., V\'azquez,R.A., Carraro, G., Baume, G.,
                                 Giorgi, E.E., Lyra, W., 2006, \mnras, 368, L77
\bibitem[Murphy \& May (1991)]{mur91} Murphy, D.C., May, J., 1991, \aap, 247, 202
\bibitem[Odenkirchen \& Brosche (1992)]{ode92} Odenkirchen, M., \& Brosche, P.\ 1992, Astronomische Nachrichten, 313, 69
\bibitem[Russeil (2003)]{rus03} Russeil, D.\ 2003, \aap, 397, 133
\bibitem[Santos-Silva \& Gregorio-Hetem (2012)]{san12} Santos-Silva, T., \& Gregorio-Hetem, J.\ 2012, \aap, 547, 107
\bibitem[Stetson (1987)]{ste87} Stetson, P.B., 1987, \pasp, 99, 191
\bibitem[Vazquez et al. (2008)]{vaz08} V\'azquez, R.A., May, J., Carraro, G., Bronfman, L.,
                                 Moitinho, A., Baume, G., 2008, \apj, 672, 930
\bibitem[Vicente et al. (2010)]{vic10} Vicente, B., Abad, C., Garz\'on, F., \& Girard, T.~M.\ 2010, \aap, 509, 62
\bibitem[Xu et al. (2013)]{xu13} Xu, Y., Li, J.~J., Reid, M.~J., et al.\ 2013, \apj, 769, 15
\bibitem[Zaxharias et al. (2012)]{zac12} Zacharias, N., Finch, C.~T., Girard, T.~M., et al.\ 2013, \aj, 145, 44
\end{thebibliography}
\end{document}